 \documentclass{article}
\usepackage{amsmath}
\usepackage{amssymb}
\usepackage{enumerate}
\usepackage{mathrsfs}
\usepackage[english]{babel}
\usepackage{fontenc}
\usepackage{graphicx}
\usepackage{graphicx}
\makeatletter
\@addtoreset{equation}{section}

\makeatother
\newtheorem{theorem}{Theorem}[section]
\newtheorem{prop}[theorem]{Proposition}

\newtheorem{remark}[theorem]{Remark}
\newenvironment{rem}{\begin{remark} \rm}{\end{remark}}

\newcommand{\bm}[1]{\mbox{\boldmath{$#1$}}}
\newcommand{\mc}[1]{{{\mathcal{#1}}}}

\newcommand{\rhob}{{\overline \rho}}

\newcommand{\oper}{{\cal H}}

\newcommand{\drho}{\rho_{{}_\Delta}}

\def\d{{\mathrm d}}

\def\oper{{{\cal H}}}

\usepackage{cancel}

\newcommand{\mb}[1]{{\mathbf{#1}}}
\newcommand{\bs}[1]{{\boldsymbol{#1}}}

\def\ddd#1#2{\displaystyle{\frac{\partial #1}{\partial #2}}}
\def\fddd#1#2{\displaystyle{\frac{\delta #1}{\delta #2}}}

\def\la{\lambda}        
\newcommand{\dsl}[1]{{\displaystyle{#1}}}
\newcommand{\RR}{{{\mathbb{R}}}}
\newcommand{\ou}{{{\overline{u}}}}
\newcommand{\dNLS}{{{$\mathrm{d}^2\mathrm{NLS}$}}}
\begin{document}
\begin{center}\medskip
{\large\bf Two-layer  interfacial flows beyond the Boussinesq \\  \medskip approximation: a Hamiltonian approach}\\ \bigskip
R. Camassa${}^1$, 
G. Falqui${}^2$, G. Ortenzi${}^2$ \\ \bigskip
 ${}^1$ Carolina Center for Interdisciplinary Applied Mathematics, Department of
  Mathematics, University of North Carolina at  
  Chapel Hill, NC 27599, USA\\ \medskip
 ${}^2$ Dipartimento di Matematica e Applicazioni, Universit\`a di Milano-Bicocca, \\ via R. Cozzi, 55, I-20125 Milano, Italy
\end{center}

\date{\today}
\begin{center}
{\bf Abstract} 
\end{center}

\noindent 
The theory of integrable systems of Hamiltonian PDEs and their near-integrable deformations is used to study evolution equations resulting from vertical-averages of the Euler system for 
two-layer stratified flows in an infinite $2D$ channel.  
The Hamiltonian structure of the averaged equations is obtained directly from that of the 
Euler equations through the process of Hamiltonian reduction. 
Long-wave asymptotics together with the Boussinesq approximation of 
neglecting the fluids' inertia is then applied to reduce the leading order vertically averaged equations to the shallow-water Airy system, and thence, in a non-trivial way, to the dispersionless 
non-linear Schr\"odinger equation. The full non-Boussinesq system for the dispersionless limit can then be viewed as  
a deformation of this well known equation. In a perturbative study of this deformation, it is shown that at first order the deformed system possesses an infinite sequence of constants of the motion, thus casting this system within the framework of completely integrable equations. The Riemann invariants of the deformed model are then constructed, and some local solutions found by hodograph-like formulae for completely integrable systems are  obtained.

\section{Introduction}
\label{intro}
\begin{figure}
\begin{center}
\includegraphics[height=5cm]{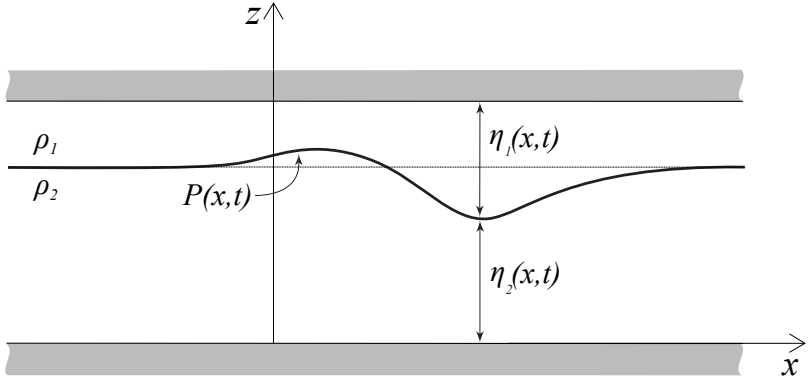}
\end{center}
\caption{Schematic of the two-layer fluid setup and relevant notation}
\label{2l-fluid-fig}
\end{figure}

Aspects of the theory of two-layer stratified flows in an infinite $2D$ channel have been the subject of intense recent studies. Layer models are widely used in a variety of geophysical applications (going back to early references such as that by Long~\cite{Lo56} in the framework of meteorology), and are of conceptual value for illustrating many fundamental properties of stratified fluid dynamics. 
A typical configuration is depicted in Figure~\ref{2l-fluid-fig}, with
an  interface 
between the two fluid representing the sharp pycnocline between superficial (``fresh'') water, 
labeled by the index $1$, and deep (``salty") water, labeled by the $2$-index. (Other   
relevant notation used throughout the paper is defined by the figure). Long internal waves in such systems were  studied in, e.g.,~\cite{CC96,CC99}, by deriving the two-layer models (including dispersive terms)
by the layer-averaging method (see e.g., \cite{Wu81}). Their dispersionless counterparts were more recently reconsidered in papers 
by Milewski, Tabak and collaborators~\cite{MTTRM04,CT6}. 
These papers were mainly interested in studying the Kelvin-Helmholtz (KH) 
instability (see also \cite{BeBr97-1}) viewed as hyperbolic vs. elliptic transition  for the resulting quasi-linear equations of motion,  
and its relation to the well-posedness of the initial value problem for these equations. In particular, for the so-called Boussinesq approximation, which in this context consists
of disregarding density differences in the inertial terms while retaining them in the buoyancy terms, the following  conditions for shear-flow stability are all equivalent:
\begin{enumerate}[i)]
 \item The ``standard" stability criterion expressed by the Richardson number  for the {\em linearized} two-layer equations (\cite{M61,BM12}) around a velocity jump.
 \item The hyperbolicity (nonlinear)  criterion for the reduced quasi-linear system of PDEs in the variables $\xi, w$ (the difference between the fluid layer thicknesses  
 and the velocity shear, respectively).
 \item A criterion given by a suitably defined Richardson-like number expressing the ratio of the kinetic energy available for the complete mixing the two layers to the potential energy barrier given
by the stratification (see e.g. \cite{MTTRM04, BCR96}).
\end{enumerate}
The equivalence between the  first two conditions holds independently of the Boussinesq approximation once the reduction to two dependent fields such as $\xi$ and $w$ is carried out. 
In contrast, in going beyond the Boussinesq approximation the equivalence of the third condition with the first two fails.
In fact, the relevant Richardson number is computed in \cite{BM12} as 
\begin{equation}\label{RiBM}
 Ri=\frac{2gr(h-r\eta_2)}{(1-r^2)\eta_2 w^2}\, , 
\end{equation}
where $r$ is half of the density difference divided by its mean, 
$\eta_2$ is the interface height, $h$ is the channel height, and $w$ is  the velocity shear at the interface.
The  ratio of the potential and kinetic energy balance mentioned in the third condition is
\begin{equation}
\frac{2gr h^2}{(h-r\eta_2) w^2} \, ,
\end{equation}
so that the two instability-related numbers need not coincide as soon as $r\neq 0$. Thus, 
the increased physical accuracy of the non-Boussinesq theory requires a modification of the stability criterion and can affect the mathematical and physical properties of the the full system such as its well-posedness and the formation of KH bellows.  

A further motivation to ``go beyond the Boussinesq approximation'' relies in some recent results obtained in \cite{CCFOP12, CCFOP13} concerning an
apparently paradoxical consequence of stratification and confinement, that is, the non-conservation of the horizontal momentum due to pressure imbalances 
at the far ends of the channel.

With this in mind, our study is organized as follows.  After a brief review of the derivation 
from the
$2D$ Euler equations of the governing equations for the evolution of the interface and suitable layer mean quantities, and a discussion of the Boussinesq approximation, we examine  the hyperbolic-elliptic transition of the 
 non-Boussinesq limit~\cite{BM12}). 
We then turn to first of the aims of the present paper, that of  fully framing the theory of two layer models within the Hamiltonian 
settings of the Euler equations. We work 
with the setting devised in \cite{Ben86}, which is specifically suited to treat heterogeneous fluids in two-dimensional domains, and does not require the introduction of 
Clebsch variables (as the original more general setting discussed in \cite{Z85, ZK97} and \cite{SW68}). 
By means of a version of the Marsden-Ratiu-Weinstein reduction procedure (see,e.g.,  \cite{MR86}), we show how the Poisson {\em structure} 
defined by Benjamin in \cite{Ben86} on the full phase space of the Euler equation 
gives rise to a well defined  ``canonical'' Poisson structure on the phase space of the reduced quasi-linear equations of the $1D$ dispersionless 
equations. 
This Poisson structure is independent of the $1D$ model, i.e., it is the same for both the Boussinesq approximation and for its 
non-Boussinesq ``deformation''; a key point for its definition is to replace, in the pair of ``coordinates'' for the $1D$ model, the velocity shear 
$w$ (used in~\cite{BM12}) with the {\em momentum shear}, a choice that is possibly the most natural in the reduction 
process of the Benjamin $2D$ Poisson structure.

Next, we discuss the Boussinesq approximation, where the velocity and momentum shear basically collapse into the same variable. From the analysis in, e.g.,~\cite{MTTRM04},\cite{BM12} the Boussinesq limit is known to be equivalent to the Airy system for long 
dispersionless waves of a single water layer over a flat bottom, under a suitable coordinate transformation  dictated 
by the structure of the characteristic velocities and the Riemann invariants of the system. In turn, this system coincides with the dispersionless defocusing NLS (``\dNLS") 
equation under the so-called Madelung  transform.
As well known, and recalled in detail in Section~\ref{B-limit}), such a system displays a lot of ``good" properties. For instance, it is
one of the few quasi-linear systems in $N>1$ fields in which the Riemann procedure can be effectively carried out (see~\cite {TY99}) 
and Whitham equations can be quite explicitly 
solved. More importantly for our purposes, this shows that the Airy/dNLS system is completely integrable infinite dimensional system, and, by means of the bi-Hamiltonian 
procedure, an explicit forms of generating functions for  the constants of the motion can be provided. As mentioned above, thanks to the fact that the Poisson structure is one and the same for the Boussinesq model as well as for its non-Boussinesq counterpart,
we can study the latter as a Hamiltonian deformation of the first. In view of its relevance for physical applications, where the non-Boussinesq deformation is scaled by the small parameter 
$r={(\rho_2-\rho_1)}/{(\rho_2+\rho_1)}$  where $\rho_1$ and $\rho_2$ are the densities of the lighter and heavier fluid respectively, we focus in particular on the 
first order deformation $O(r)$.  
 We show that the first order deformed system retains the property of being completely integrable, that is, we explicitly  prove 
that one of the three families of mutually commuting integrals of motion for the Boussinesq-Airy system can be deformed to integrals of motion of the first order deformed Hamiltonian system. 
These integrals are conjectured to be in involution,  and solutions for families of initial data, following hodograph-like formulas for completely integrable systems, can be provided. 
\section{The layer-averaged equations of motion}\label{eqm-sect}
We briefly review the derivation of the layer-averaged equations from the corresponding Euler system
for a two-layer  incompressible Euler fluid in an infinite channel (see, e.g.,  \cite{CC99}).

Motions of typical wavelength $L$ were considered, under the assumptions that the ratios
\begin{equation}\label{epsi}
\epsilon=\frac{h}{L}\simeq\frac{\eta_i}{L}
\end{equation}
can be considered small, where $h$ is the total height of the channel, 
while $\eta_1$ (resp. $\eta_2$) is the thickness 
of the upper (resp. lower) fluid. The densities of the two fluids are denoted by $\rho_1$ and $\rho_2$ ($\rho_2\ge\rho_1$).  

Under assumption~(\ref{epsi}) the ratio of vertical and horizontal velocities scales 
as $\epsilon$ as well, and by using the layer-averaging method as described in~\cite{Wu81},  
the $2+1$-dimensional  Euler system together with the incompressibility of each layer, \begin{equation}
\label{Eceq}
\left\{
\begin{array}{l}
\rho_t+\rho_xu+\rho_zw=0\\
u_t+uu_x+wu_z=-{P_x}/{\rho}\\
w_t+uw_x+ww_z=-{P_z}/{\rho}-g\\
u_x+w_z=0 \, , 
\end{array}
\right.
\end{equation}
under some further assumptions (see~\cite{CC99}) reduce to the $1+1$ dimensional equations  
\begin{equation}
\label{2layer}
 \begin{split}
 & {\eta_i}_t+(\ou_i\eta_i)_x=0, \, \quad i=1,2 \\ &
 {\ou_1}_t+{\ou_1}{\ou_1}_x -g {\eta_1}_x + \frac{P_x}{\rho_1} + D_1 =0, \\
 &  {\ou_2}_t+{\ou_2}{\ou_2}_x +g {\eta_2}_x + \frac{P_x}{\rho_2} + D_2 =0
 \\ &\eta_1+\eta_2=h, \qquad (\eta_1 \ou_1 + \eta_2 \ou_2)_x=0
 \end{split}
\end{equation}
where   $D_i= \frac{1}{3 \eta_i} [\eta_i^3 ({\ou_i}_xt +{\ou_i}{u_i}_xx-({\ou_i}_x)^2)]_x  + \dots$ are dispersive terms.
Here $\ou_{1,2}$ are the layer-mean velocities, defined as
\[
\ou_1(x,t):=\frac{1}{\eta_1(x,t)}\int_{h-\eta_1(x,t)}^{h} u_1(x,z,t)\d z, \quad \ou_2(x,t)=\frac1{\eta_2(x,t)}\int_0^{\eta_2(x,t)} u_2(x,z,t) \d z\, ,
\]
 and $P(x,t)$ is the interfacial pressure. 
We shall always assume, consistently with (\ref{epsi}), that the interface 
$\eta(x,t)\equiv \eta_2(x,t)$  nowhere and never touches the boundary, that is,
\begin{equation}
0<\eta(x,t)<h, \text{ and   } \eta_1\simeq \eta_2.
\end{equation}
The constraints in the last line of (\ref{2layer}) reduce the full system to evolution equations for just two fields. Indeed, under the assumption of vanishing horizontal velocities at the far end of the channel (that is, for $|x|\to\infty$), the constraints can be algebraically solved say for 
$(\eta_2, \ou_2)$ as
\begin{equation}
\label{solvconstr}
\eta_1=h-\eta_2,\quad \ou_1=-\frac{\eta_2}{h-\eta_2}\ou_2.
\end{equation}
The constrained equations of motion can be obtained retaining the volume conservation of the lower fluid (
$ \eta_{2,t}+(\ou_2\eta_2)_x=0$ and 
eliminating $P_x$ from the second and third line of (\ref{2layer}). 

In what follows, we shall choose as reduced coordinates the relative thickness $\xi=\eta_2-\eta_1=2\eta_2-h$ and the {\em momentum shear}
\begin{equation}\label{sigdef}
\sigma=\rho_2\ou_2-\rho_1\ou_1.
\end{equation}
In these variables the resulting equations read
\begin{equation}\label{e-m-full}
\left\{
\begin{array}{cl}\medskip
\dsl{ \xi_t= }&\dsl{-\left( {
\frac { \left( h^2-\xi^2 \right) \sigma} { (\rho_2+\rho_1) h + (\rho_2-\rho_1) \xi}} \right)_x}
 \\
\dsl{\bar{\sigma}_t}=
&\dsl{-\left( \frac{\rho_2 (h-\xi)^2 -\rho_1 (h+\xi)^2 }{2(  (\rho_2+\rho_1) h + (\rho_2-\rho_1) \xi )^2}\, \sigma^2
+\frac{g(\rho_2-\rho_1)}{2}\xi \right)_x}
\end{array}
\right.
\end{equation} 

\begin{rem}
As it might seem more natural from a physical viewpoint, the choice made in~\cite{BM12} is 
to complement to relative thickness $\xi$ with  the {\em velocity} shear $w=\ou_2-\ou_1$.  The velocity shear  and the momentum shear   $\sigma$ are simply related by 
\begin{equation}\label{w2sigma}
\sigma= \frac{(\rho_2+\rho_1)w}{2}\left(1 -\frac{\rho_2-\rho_1}{\rho_2+\rho_1}\frac {\xi}{h} \right) \, .
\end{equation}
The reasons behind our choice of this second dependent variable will be fully motivated and discussed in Section \ref{Sect-1}.
\end{rem}

The so-called {Boussinesq approximation}, widely used in the theory of  in the field of buoyancy-driven flow, consists in neglecting 
small density differences except for the gravity terms. As well-known, this can be viewed 
as the double scaling limit obtained by setting 
\begin{equation}\label{rhodeltadef}
 \drho \equiv \rho_2-\rho_1
\end{equation} 
and considering
\begin{equation}\label{douscalim}
\drho \to 0,\quad  g \to \infty, \quad \text{with} \quad \drho\, g\sim O(1).
\end{equation} 
In the Boussinesq approximation the momentum shear $\bar\sigma$ is simply a multiple of the velocity shear,
\begin{equation}\label{sigbl}
 \bar\sigma\to \bar\rho(u_2-u_1), 
\end{equation} 
and the  equations of motion become
\begin{equation}\label{e-m-Boussi}
\left\{
\begin{array}{cl}\medskip
\dsl{ \xi_t= }
&\dsl{- {\frac {\big(( h^2-\xi^2 \big) \sigma \big)_x} { 2 \bar{\rho} h }} }
 \\
\dsl{\bar{\sigma}_t}=
&\dsl{-\left( -\frac{ \xi \sigma^2}{ 2\bar{\rho} h  }\, 
+\frac{g\drho}{2}\xi \right)_x}
\end{array}
\right.
\end{equation} 
\begin{rem}
The limit $r\to0$  with $g$ fixed (finite) is drastically different, and  the properties of such a model are briefly discussed in Appendix \ref{naif}.
\end{rem}

\subsection{Characteristic velocities and KH instability}\label{KH}
 The study of the hyperbolic-elliptic transition for the Boussinesq system (\ref{e-m-full}) has been considered (by using different coordinates)  in \cite{BM12}. We report some results obtained in
that paper and comment on them by first rewriting them in terms of our choice of coordinates. 
The hyperbolicity region is  defined by
\begin{equation}
\label{hr}
  |\sigma| < \sqrt{\frac{ g (\rho_2-\rho_1)( \rho_1 (h+\xi) + \rho_2 (h-\xi))^{3}}{8 h^2 \rho_1\rho_2}} = 
  \bar{\rho}\sqrt{\frac{ 2 g rh  (1 - r \xi/h  )^{3}}{1-r^2}}:=\sigma_b \,
 , \quad |\xi|<h. 
\end{equation}
Notice that the only condition on  the $\xi$ variable is that 
the interface does not touch the channel boundaries, i.e., $|\xi| < h$.

\begin{figure}[htb]
\centering
\includegraphics[width=8cm, height=4.5cm]{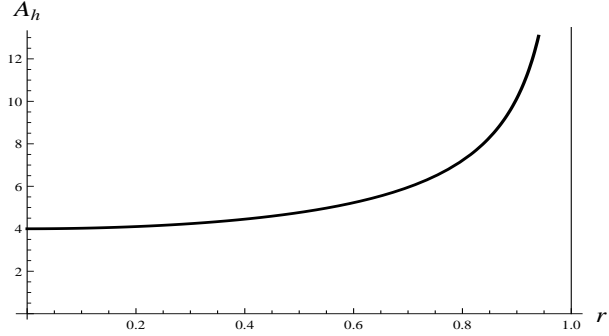}
\caption{Dependence of the measure $A_h$, in units of $\bar{\rho} h \sqrt{2gh}$, of the hyperbolicity region as function of the inertia parameter $r$. }
\label{hypreg-area}
\end{figure}

\begin{figure}
\centering
\includegraphics[width=10cm, height=5cm]{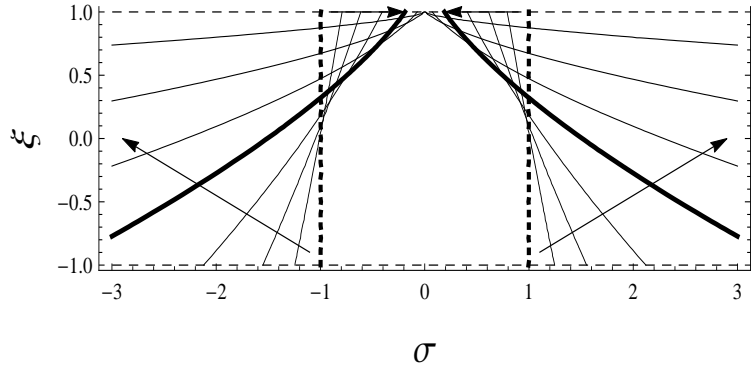}
\caption{The hyperbolicity region for increasing inertia parameter $r$. Arrows indicate the direction of deformation of this region's boundary as $r$ grows from the Boussinesq case $r=0$ to values $r>0$.  The dashed lines correspond to the locations of the top and the bottom of the 2D channel.
The vertical dotted thick lines are the boundary of the hyperbolicity region for the Boussinesq limit. 
The continuous thick line corresponds to $r=.75$. The variables $\xi$ and $\sigma$ are measured 
in units of $h$ and $\bar{\rho}\sqrt{2\tilde{g}h}$, respectively (note that we have switched the order of axes to have $\xi$ on the ordinate for this figure).}
\label{hypreg}
\end{figure}

The parameter $r$ affects substantially the structure of the hyperbolicity region. Indeed 
the area $A_h$ of such a region  is the monotonically increasing function of $r$ 
\begin{equation}
 A_h= \rhob  \sqrt{2 g r h^3} \frac{4 \left((1+r)^{5/2}-(1-r)^{5/2}\right)}{5 r \sqrt{1-r^2}}\, ,
\end{equation}
whose graph is in Figure~\ref{hypreg-area}. 
Notice that $A_h$ goes to a finite quantity (in the units we  are using, the limit is $A_{h,0}=4$) when $\rho_2 \to \rho_1$, and grows indefinitely (as  $(1-r)^{-\frac12}$) when $r\to 1$, the limiting case of an air-water system. The fact that the area $A_h$ grows monotonically with $r$ should be expected on the basis of the stabilizing effects of stratification.
In Figure~\ref{hypreg} we depict the explicit form of the hyperbolicity domain in the Boussinesq expansion.

As remarked in \cite{BM12}, the case $r=0$  corresponds to the Boussinesq approximation. 
The domain 
of hyperbolicity is finite ($|\sigma|=\bar{\rho}\sqrt{2\tilde{g}h}$), 
the hyperbolic-elliptic transition is forbidden because the simple waves are tangent to the sonic line.
In the opposite limit ($r=1$) the system becomes air-water like and  
the hyperbolicity region fills the entire domain of the variables. 
Simple waves in general satisfy the equation 
\begin{equation}
\label{SW}
 \frac{\d \sigma}{\d \xi}= \rhob \,\,  \sqrt{\frac{2 g r (h-\xi  r)^3-h^2 \left(1-r^2\right) \sigma ^2}{\left(h^2-\xi ^2\right) (h-\xi  r)^2}}
 .
\end{equation}
The tangent lines to  simple waves at the boundaries of the hyperbolic region are always independent of $r$
\begin{equation}
 \frac{\d \sigma}{\d \xi} \Big{|}_{|\xi|=h}=\infty, \qquad \qquad 
 \frac{\d \sigma}{\d \xi} \Big{|}_{|\sigma|=\sigma_b}=0 \, .
\end{equation}
On the other hand the tangent $t_s$ to the sonic line $s$  (the line in the $(\xi,\sigma)$-plane 
where the two characteristic velocities coincide) 
at $|\sigma|=\sigma_b$ is
\begin{equation}
\label{tHEl}
 t_s=({-3 r \sqrt{\tilde{g}(h-\xi  r)}},{ \sqrt{2(1-r^2)}}).
\end{equation}
Therefore the only two values of $r$ which prevent hyperbolic-elliptic transition are $r=0$ and $r=1$.\par
In  table \ref{tabsmall-bigr} we resume the behavior the hyperbolic region for small and big $r$ in the Boussinesq expansion. In the case of small $r$ the hyperbolicity region is a trapezoid and the area is constant at order $O(r)$.  In the case of $r \sim O(1)$ the 
ellipticity region restricts, for not too big $\sigma$, to a strip near to $\xi=h$. Indeed, if the interface
remains far from the upper lid, the system behaves as a free surface fluid, while if the interface goes near to the upper lid 
the incompressibility of upper light fluid  introduces an instability in the system.\\
\begin{table}[h]
\begin{center}
\begin{tabular}{|c|c|c|c|c|}
\hline &&&& \\
Value of $r$ & Hyperbolicity region & $A_h$ & tangent to the   & tangent to the\\ & & & sonic curve $s$ & simple wave $(\dot{\xi},\dot{\sigma})$ at $s$ 
\\&&&&\\
\hline
&&&&\\
$r \sim o(1)$ & $|\sigma| < 1-\frac{3 r \xi}{2}$ & $4$ &  $( 2 ,3r )$ &  $(1,0)$ \\
&&&&\\
\hline
&&&&\\
$r \sim O(1)$ & $|\sigma| < -\frac{(1-\xi )^{3/2}}{\sqrt{2-2 r}}$ & $\frac{16}{5 \sqrt{1-r}}$ & $
({2 \sqrt{2-2 r}},{3 \sqrt{1-\xi }})$ &  $(1,0)$ \\
&&&& \\ 
\hline
\end{tabular}
\end{center}
\caption{The $r$-expansion around the Boussinesq limit for the hyperbolicity domain of measure $A_h$. Coordinates  ($\xi,\sigma$) are measured, respectively, in units of 
$h$ and $ \rhob \sqrt{2 \tilde{g} h}$. }
\label{tabsmall-bigr}
\end{table}

\section{The $2D$ Benjamin model for heterogeneous fluids in a channel}\label{Sect-1}
Benjamin \cite{Ben86} proposed and discussed a  set-up for the Hamiltonian formulation of the incompressible stratified Euler system, also known as the Boussinesq model (not to be confused with its namesake approximation, which only refers to neglecting density variations in the fluid's inertia). We hereafter summarize, for the reader's convenience, his results.
 
The evolution 
of a perfect inviscid incompressible but heterogeneous fluid in 2D, subject to gravity, 
is usually described by the variables $\rho(x,z,t), \mb{u}=(u(x,z,t), w(x,z,t))$, governed by the Euler equations (\ref{Eceq}).
Benjamin's idea was to consider, as  basic variables, the density $\rho$ together with the ``weighted vorticity" $\sigma$ defined by
\begin{equation}
\label{sigmadef}
\sigma=\nabla\times (\rho\bs{u})=(\rho w)_x-(\rho u)_z. 
\end{equation}
The equations of motion for these two fields, ensuing from (\ref{Eceq}), are 
\begin{equation}
\label{eqsr}
\begin{array}{l}
\rho_t+u\rho_x+w\rho_z =0\\
\sigma_t+u\sigma_x +w\sigma_z +\rho_x\big(gz-\frac12(u^2+w^2)\big)_z+\frac12\rho_z\big(u^2+w^2\big)_x=0\ .
\end{array}
\end{equation}
They can be written in the form
\begin{equation}
 \label{heq}
{\rho_t}=-\left[\rho,  \dsl{\fddd{H}{\sigma}}\right] \, , \qquad 
\sigma_t= -\left[\rho,  \dsl{\fddd{H}{\rho}}\right]-\left[\sigma, \dsl{\fddd{H}{\sigma}}\right] \, ,
\end{equation}
where, by definition, $[A, B] \equiv A_xB_z-A_zB_x$, and the  functional 
\begin{equation}
\label{ham-ben}
H= \dsl{\int_\mathcal{D} \rho\left(\frac12 |\bs{u} |^2+g z\right)\,{\rm d}x\,{\rm d}z, }
\end{equation}
is simply given by the sum of the kinetic and potential energy, $\mathcal{D}$ being the fluid domain.
The most relevant feature of this coordinate choice is that $(\rho,\sigma)$ are physical variables.
Their use, albeit confined to the 2D case, allows one to avoid  the introduction of Clebsch 
variables (and the corresponding subtleties associated with gauge invariance of the Clebsch potentials) 
needed in the Hamiltonian formulation general case (see \cite{Z85}).

Actually, in Benjamin's  formalism, 
the Hamiltonian $H$ is  written in terms of the stream function $\psi$
related to the variables  $(\rho, \sigma)$ via
\begin{equation}
\label{Sigmadef}
\sigma=(\rho w)_x-(\rho u)_z=-(\rho\psi_x)_x-(\rho\psi_z)_z=
-\rho\nabla^2 \psi-\nabla\rho\cdot \nabla\psi.
\end{equation}
More precisely, once $\rho$ and $\sigma$ are given, $\psi$ is the unique solution of (\ref{Sigmadef}) vanishing on the plates, so that $H$ turns out to be a functional of $\rho$ and $\sigma$ only.
As shown by Benjamin, equations (\ref{heq}) are actually a Hamiltonian system with respect to a 
non-canonical Hamiltonian structure. 
This means that equations (\ref{eqsr}) can be written as
\[
 \rho_t=\{\rho, H\}
,\qquad \sigma_t=\{\sigma, H\}
\]
for the Poisson brackets defined by the Hamiltonian operator
\begin{equation}\label{B-pb}
J_B=-
\left(\begin{array}{cc}
       0 & \rho_x \partial_z -\rho_z \partial_x \\ 
       \rho_x \partial_z -\rho_z \partial_x & \sigma_x \partial_z -\sigma_z \partial_x
      \end{array}
\right).
\end{equation}
\subsection{The Hamiltonian Reduction}\label{channel-averaging}
We now discuss how a simple averaging process can be given a Hamiltonian structure, 
well suited to the discussion of the constrained equation in which  
our set of reduced coordinates naturally appears.
In particular, we can induce, at order $o(\epsilon)$, where $\epsilon=h/L$ as before, 
a Poisson bracket on the reduced fields from the full 2D Benjamin's structure~(\ref{B-pb}).

By means of the Dirac and Heaviside generalized functions, a two-layer fluid with 
constant density $\rho_i$  and velocity components $u_i(x,z),w_i(x,z)$   
for the upper layer $i=1$ and lower layer $i=2$, respectively,  can be described 
by global density and velocity variables defined as
\begin{equation}\label{no1}
\begin{split}
&\rho(x,z)=\rho_2+(\rho_1-\rho_2)\theta(z-\eta(x))\\
&u(x,z)=u_2(x,z)+(u_1(x,z)-u_2(x,z))\theta(z-\eta(x))\\
&w(x,z)=w_2(x,z)+(w_1(x,z)-w_2(x,z))\theta(z-\eta(x))\, .
\end{split}
\end{equation}
Since time is merely a parameter in the description of phase spaces and mappings, we suppress time dependence for ease of notation in the following .

The fluid velocity vector $\bm{u}=(u,w)$ is assumed to be
smooth except at the interface $z=\eta(x)$ where it may have finite jumps, 
subject to the continuity of the normal component, where the density discontinuity is 
located. 

As can be easily checked, the two momentum components are 
\[
\rho u= \rho_2 u_2(x,z)+(\rho_1 u_1(x,z)-\rho_2 u_2(x,z))\theta(z-\eta(x))\, , 
\]
and 
\[
\rho w= \rho_2 w_2(x,z)+(\rho_1 w_1(x,z)-\rho_2 w_2(x,z))\theta(z-\eta(x)) \, . 
\]
Hence, by the standard rules of differentiation of generalized functions, 
$\theta'(\cdot)\equiv\delta(\cdot)$,
the terms in the weighted vorticity $\sigma=(\rho w)_x-(\rho u)_z$ become
\begin{equation}
\label{no2a}
\begin{split}
(\rho w)_x&=\rho_2 w_{2,x}+(\rho_1 w_{1,x}-\rho_2 w_{2,x})\theta(z-\eta(x))+\\ &-\eta_x(\rho_1 w_1(x,z)-\rho_2 w_2(x,z))\delta(z-\eta(x)),\end{split}
\end{equation}
and
\begin{equation}
\label{no2b}\begin{split}
(\rho u)_z&=\rho_2 u_{2,z}+(\rho_1 u_{1,z}-\rho_2 u_{2,z})\theta(z-\eta(x))+\\
&+(\rho_1 u_1(x,z)-\rho_2 u_2(x,z))\delta(z-\eta(x))  \, ,
\end{split} 
\end{equation}
so that
\begin{equation}\label{no-s}
\begin{split}
\sigma=&\rho_2( w_{2,x}-u_{2,z})+\big(\rho_1 (w_{1,x}-u_{1,z})-\rho_2( w_{2,x}-u_{2,z})\theta(z-\eta(x))+\\
&-\big(\rho_1 u_1(x,z)-\rho_2 u_2(x,z)+\eta_x(\rho_1 w_1(x,z)-\rho_2 w_2(x,z))\big)\delta(z-\eta(x))\, .
\end{split}
\end{equation}
If the motion in each layer is assumed to be irrotational the first line in this expression for $\sigma(x,z)$ is identically zero, and we are left with the interface localized form (a ``momentum vortex sheet")
\begin{equation}\label{no=s2}
\sigma=\big(\rho_2 u_2(x,z)-\rho_1 u_1(x,z)+\eta_x(\rho_2 w_2(x,z)-\rho_1 w_1(x,z))\big)\delta(z-\eta(x)).
\end{equation}
%
In these coordinates 
the projection map 2D $\to$ 1D is easily established. 
Indeed, 
\begin{eqnarray}
h\, \overline{\sigma}\equiv \int_0^h \sigma(x,z)\, dz&=&
\rho_2 u_2(x,\eta)-\rho_2 u_1(x,\eta)+\eta_x(\rho_2 w_2(x,\eta)-\rho_1 w_1(x,\eta)) 
\nonumber
\\ 
&=&\rho_2 {\tilde u}_2(x)-\rho_2 {\tilde u}_1(x)+\eta_x(\rho_2 {\tilde w}_2(x)-\rho_1 {\tilde w}_1(x)) 
\label{intersi}
\end{eqnarray}
where we have introduced the notation $\tilde{\bm{v}}$ for the velocity at the interface. Thus, the averaged 
$\overline \sigma$ reduces to the weighted tangential momentum jump at the interface. 

For long wave dynamics, the components of velocity
at the interface for each fluid can be expressed as an asymptotic expansion of layer-averaged horizontal velocities $\overline u_i$'s (see, e.g., \cite{CC96, Wh74})
in terms of the small parameter $\epsilon$.
The right hand side of equation~ (\ref{intersi}) can then be written as 
\begin{equation}\label{dopola3.15}
\rho_2 \overline{u}_2(x)-\rho_1 \overline{u}_1(x)+O(\epsilon), 
\end{equation}
that is, in the long wave regime the averaged weighted vorticity reduces to the  weighted horizontal momentum jump (a localized shear) at the interface.
We define the reducing map as
\begin{equation}\label{mappazza}
 \eta=\dsl{\frac{1}\drho}
 \int_0^h (\rho(x,z)-\rho_1)\, \d z \, , \qquad
\overline{\sigma}=\frac1h\int_0^h \sigma(x,z) \d z \, 
\end{equation}
(the first of these relation being easily obtained from the first of equations (\ref{no1}) and from  equation~(\ref{no=s2})),
and we can compute the  reduced Poisson tensor can be computed by means of the standard ``pull-push" formula (see, e.g., \cite{MR86}). Let us denote by $M^{(1)}$ 
the manifold of the ``averaged" quantities $(\overline{\sigma},\eta)$ and by $M^{(2)}$ the manifold of 
the 2D quantities, parametrized by $(\rho(x,z), \sigma(x,z))$. Inside $M^{(2)}$ we consider the surface
\begin{equation}\label{intermfld}
\mathcal{I}:=
\{\rho(x,z)=\rho_2+
\drho\theta(z-\eta(x)), 
\sigma(x,z)=(\rho_1 \bar{u}_1(x)-\rho_2 
\bar{u}_2(x))\delta(z-\eta(x))\}\, .
\end{equation}
If $(\alpha_\eta(x), \alpha_{\bar{\sigma}}(x))$ is a 1-form on $M^{(1)}$, we lift this, under the map (\ref{mappazza}), to obtain 
 \begin{equation}
 \label{lift}
 \left(\frac1\drho
 \alpha_\eta(x),\> \frac1{h}\alpha_{\bar{\sigma}}(x)\right)
 \end{equation} 
 (it is independent on the vertical coordinate $z$).
 
 Applying the Poisson tensor (\ref{B-pb}) evaluated on $\mathcal{I}$ to this covector we get (notice that terms with $\partial_z$ can be dropped)
 \begin{equation}\label{PS}
 \left( 
 \begin{array}{c}
 \dot\rho(x,z)\\
\dot\sigma(x,z)
 \end{array}\right)
 =-
 \left(\begin{array}{cc} 0&\drho
 \delta(z-\eta(x))\partial_x\\
 \drho\delta(z-\eta(x))\partial_x&{h\bar{\sigma}(x)\delta_z(z-\eta(x))\partial_x}
       \end{array}\right)
       \left(\begin{array}{c} \dsl{\alpha_\eta(x)/{\drho}}
       \\
       \alpha_{\bar{\sigma}}(x)/h
       \end{array}\right) 
 \end{equation} 
This gives
\begin{equation}\label{PS2}
\left( 
 \begin{array}{c}\bigskip
 \dot\rho(x,z)\\ 
\dot\sigma(x,z)
 \end{array}\right)
 =-\left( 
 \begin{array}{c}\medskip
 \drho\delta(z-\eta(x))\left( 
 \frac1{h}\alpha_{\bar{\sigma}}(x)\right)_x
 \\
\drho\delta(z-\eta(x))\left(\dsl{\frac1{\drho}}
\alpha_\eta(x)\right)_x+\dsl{\frac{\bar{\sigma}(x)}{h}}\delta_z(z-\eta(x))
\big(\alpha_{\bar{\sigma}}(x)\big) 
 \end{array}\right)\, .
 \end{equation} 
 Pushing this vector to $M^1$ via the tangent map to (\ref{mappazza})   yields the vector
\[
(\dot{\eta},\dot{\bar{\sigma}})=\left(-\frac{1}{h}\partial_x(\alpha_{\bar{\sigma}})\ \, 
-\frac{1}{h}\partial_x(\alpha_{\eta})\right),
\]
owing to the fact that $\int_0^h \delta'(z-\eta(x))\,\d z =0$.
Hence, we have shown that the reduction of the Benjamin Poisson tensor (\ref{B-pb}) is given by the  tensor
\begin{equation}\label{Pred}
J_{\rm red}:=-\frac1{h}\left(\begin{array}{cc}
0&\partial_x\\
\partial_x&0\end{array}
\right)
 \end{equation} 
\begin{rem}
This structure was termed {\em canonical} in the recent literature (see, e.g., \cite{Du08}) since it corresponds to that of the non-linear wave equation in $1+1$ dimensions
\[
u_{tt}=F''(u) u_{xx}
\]
with ``canonical" Hamiltonian functional 
$$\mc{H}=\frac12\int_{\mathbb{R}} (u_t^2+F(u)) \, dx \, ,  
$$ 
for some twice differentiable function $F(u)$.
This is in contrast with the well-known~\cite{Z85,ZK97} standard canonical symplectic formulation for the Euler equations by means of Clebsch variables.
To avoid possible misunderstandings, we shall refer to the structure (\ref{Pred}) as the {\em Darboux} structure, and to the ``canonical" variables such as $(\overline{\sigma},\eta)$ as Darboux variables.
  \end{rem}


\subsection{The reduced Hamiltonian}
The next step is construct the reduced Hamiltonian; this is done by evaluating the full $2D$ Hamiltonian (\ref{ham-ben}) 
on the manifold $\mathcal{I}$. The potential term 
\[
U=\int_{\mathcal{S}} \rho(x,z) \, g\, z\, dx dz 
 \]
is readily seen to reduce to 
\begin{equation}\label{U-red}
U_{\rm red}=\int_{\mathbb{R}} \frac{g}2\, (\rho_2-\rho_1)\eta^2 \, dx \, .
\end{equation}
The kinetic term is subtler. The main idea from long wave-asymptotic is that, at order $O(\epsilon^2))$, we can disregard  the term $w^2(x,z)$, and trade the
horizontal Euler velocities \
with the layer-averaged ones. That is, we  write the kinetic energy density as
\begin{equation}\label{Tred}\begin{split}
T=&\frac12\rho_2 ({u_2}^2(x,z)+{w_2}^2(x,z)\theta(\eta-z)+ \frac12\rho_1 ({u_1}^2(x,z)+{w_1}^2(x,z)\theta(z-\eta)\\ &=
\frac12\rho_2 ({\bar{u}_2(x)}^2\theta(\eta(x)-z)+\frac12\rho_1 ({\bar{u}_1(x)}^2\theta(z-\eta(x))+O(\epsilon^2),\end{split}
 \end{equation} 
and perform the integration along $z$ to get
\begin{equation}
 T_{\rm red}=\frac12\left(\rho_2 \eta \bar{u_2}^2+\rho_1 (h-\eta) \bar{u_1}^2\right)=\frac12\frac{\eta}{h-\eta}{\bar{u}_2}^2\cdot (\rho_2+\frac{\eta}{h-\eta}\rho_1).
\end{equation} 
where  the dynamical constraint
\[
 \bar{u}_1=-\frac{\eta}{h-\eta} \bar{u}_2
\]
has been taken  into account in the last equality. Introducing   the (lineal) mass density 
 $\varphi$ as
\begin{equation}\label{psidef}
 \varphi:=\rho_2\eta_1+\rho_1\eta_2\equiv \rho_2 h-(\rho_2-\rho_1)\eta\, , 
\end{equation} 
and taking the constraints into account, the link between our variable $\bar{\sigma}=\rho_2 \bar{u}_2-\rho_1\bar{u}_1$ and $\bar{u_2}$ is 
\begin{equation}
 \bar{\sigma}=\frac1{h-\eta} \varphi\ \bar{u}_2 \, ,
\end{equation} 
so that we can write for the reduced kinetic energy density $ T_{red}$
\begin{equation}\label{Tred-2}
 T_{red}=\frac12 \frac{\eta (h-\eta)}{\varphi}\, \bar{\sigma}^2.
\end{equation} 
Hence, together with the reduced potential energy~(\ref{U-red}), the reduced Hamiltonian functional  is
\begin{equation}\label{Hred}
{\mathcal H}_{\rm red}[\eta,\bar\sigma]=\int_\RR\left(\frac12 \left(\frac{\eta (h-\eta)}{\varphi}\, \bar{\sigma}^2\right)+\frac{g}2(\rho_2-\rho_1)\, \eta^2\right)\, \d x\, ,  
\end{equation} 
and the equations of motion are therefore
\begin{equation}\label{e-m-full-en}
\left\{
\begin{array}{cl}\medskip
\dsl{ \eta_t=}&\dsl{-\frac1{h}\partial_x\left({\frac {\eta\, \left( h-\eta \right) \sigma}{\rho_{{2}}h- \left( \rho_
{{2}}-\rho_{{1}} \right) \eta}}\right)}
 \\
\dsl{\bar{\sigma}_t}=&\dsl{-\frac1{h}\partial_x\left(\frac12 \frac{( \rho_{{2}}-\rho_{{1}}) {\eta}^{2}-2\,\eta\,h\rho_{{2}}+
{h}^{2}\rho_{{2}}}{(\rho_{{2}}h- ( \rho_{{2}}-\rho_{{1}}) \eta)^2}\, \sigma^2
+g(\rho_2-\rho_1)\eta\right)}
\end{array}
\right.
\end{equation} 

\vspace{0.3cm}
 \begin{rem}
 A few remarks are in order.
 \begin{enumerate}
 \item The reduction process clearly shows that the most appropriate variables to be used in the Hamiltonian 
 reduction process are the $(\xi,\bar{\sigma})$ variables. Indeed using the variables $(\xi,w)$ of \cite{MTTRM04, BM12} the Hamiltonian structure of the  reduced equations away from $r=0$ is not apparent and  can be shown to depend on the small parameter $r$ (as well as on the Hamiltonian). 
 Since, in our variables,   the Poisson tensor does not depend on $r$, 
 the  expansion in small $r$ of the system we are considering  can be simply obtained by the correspondent expansion of the Hamiltonian functional, keeping one and the  same  Poisson structure for all orders.
 \item
 A natural symmetry of the system is the translational invariance which is related, thanks to Noether's theorem, to the conserved quantity
 \begin{equation}
  \mu=\left(1+ \frac{\eta_1 / \rho_1}{(h-\eta_1) / \rho_2} \right) \rho_1 \eta_1 \overline{u}_1\, .
 \end{equation}
 The Hamiltonian representation (in the dispersionless limit) in the variables $\mu$ and $\zeta=-\eta_2$ has been studied in \cite{CFOP14-2}.
In this setting the Poisson structure becomes a linear Lie-Poisson structure
\begin{equation}
J_{LP}(\mu,\eta)=-\left(  
\begin{array}{ccc}
       \mu \partial +\partial \mu         && \zeta \partial \\
\partial \zeta && 0
                \end{array} 
\right),
\end{equation}
(with the standard operator notation $\partial$ for differentiation with respect to $x$) and the Hamiltonian density is 
\begin{equation}\label{HamRed}
 \mathcal{H}_{LP}(\mu,\zeta)=-\frac{\zeta \mu^2}{2 (h+\zeta)(\rho_2(h+\zeta) - \rho_1 \zeta)} 
 +\frac{g}{2}\left(\rho_2\zeta^2-\rho_2(h+\zeta)^2\right).
\end{equation}
 
\item The higher order terms in the $r$ expansion involve nonlinear differential operators acting on the 
layer-averaged horizontal velocities, and contribute dispersion to the resulting motion equation, as it is apparent from the coupled  Green-Naghdi equations (\ref{2layer}). However,  
it should be stressed that they  affect only  the reduction of the Hamiltonian functional, and not the reduction of the Benjamin Poisson brackets $J_B$ of equation \ref{B-pb}. Indeed, all the technical computations done from 
(\ref{lift}) to (\ref{Pred}) to arrive at the reduced Poisson tensor hold {\em verbatim} if we forget about the "dispersionless limit" of the variable $\bar{\sigma}$ of equation (\ref{dopola3.15}) retaining the definition (\ref{intersi}), i.e., 
\begin{equation}\label{dfintersi}
h\, \overline{\sigma}(x)= \left(\rho_2 u_2(x,\eta)-\rho_2 u_1(x,\eta)+\eta_x(\rho_2 w_2(x,\eta)-\rho_1 w_1(x,\eta))\right),
\end{equation} 
and correspondingly modify the definition of the "manifold" $\mc{I}$ in equation (\ref{intermfld}).
\item The Hamiltonian nature of the $1D$ two-layer equations was previously described in the literature
(see, e.g,  \cite{BeBr97-1,CG00,LDZ99} and the more recent \cite{DIT15}). However, while treating the general case, all these previous approaches focus on the variational setting of these equations, and define suitable variational $1D$ principles  from their $2D$ counterparts (thus reducing the Hamiltonian {\em equations}). The setting which we have just implemented, as already remarked,  is different and has a more geometrical flavor in that it deals first with the reduction of the Poisson brackets, and them with the definition of the reduced Hamiltonian, in the spirit of the geometrical Hamiltonian reduction scheme.

\item A further Hamiltonian form of our system is obtained by noticing that, once the bulk vorticity is assumed to vanish, velocity potentials $\phi_i$ can be defined in each of the two layers. Then, by introducing, as in\cite{DIT15}, the Clebsch-like variable
\begin{equation}\label{Pot}
\Phi(x)=(\rho_2 \phi_2(x,\eta(x))-\rho_1\phi_1(x,\eta(x))),
\end{equation} 
from (\ref{dfintersi}) we get $h\bar{\sigma}(x)=\partial_x(\Phi(x))$. Under (the inverse of) this transformation our reduced Poisson tensor (\ref{Pred}) turns  into the standard symplectic one, yielding the motion equations in the standard form
\begin{equation}
\label{vareq}
\xi_t=\frac{\delta H}{\delta \Phi},\quad \Phi_t=-\frac{\delta H}{\delta \xi},
\end{equation}
for a suitable $H$. These variables are the $1D$ counterparts of the classical Zakharov setting of the incompressible Euler equations~\cite{Z85, ZK97}. The equivalence between the Zakharov and the Benjamin setting for  Euler  incompressible heterogeneous 2D fluids has been discussed in \cite{CCFOP13} and \cite{CFOP14-2}.
\end{enumerate}

 \end{rem}
\subsection{The Boussinesq limit}
In the Hamiltonian theory, the Boussinesq approximation can be performed at the level of the 
Hamiltonian density, in view of the fact that  the reduced 
Poisson tensor does not depend on the densities $\rho_i$. 
Recall that the Darboux variable $\bar\sigma$ goes into $\bar\rho(u_2-u_1)$,  
 $\varphi$ limits to $ \bar{\rho} h$, while $\widetilde{g}=\drho g$ is of order $1$.

Performing this ``limit'', the 
Hamitonian density of the Boussinesq approximation becomes (see (\ref{Hred}))
\begin{equation}\label{HredB}
H_{{\rm red},B}(\eta,\bar{\sigma})=\frac12\left(\frac{\eta (h-\eta)}{\bar\rho h}\, \bar{\sigma}^2\right)+\frac{g}2(\drho)\eta^2,
\end{equation} 
and the equations of motion get simplified as
\begin{equation}\label{e-m-bou}
\left\{
\begin{array}{cl}\medskip
\dsl{ \eta_t=}&\dsl{-\left( {\frac{\eta\, \left( h-\eta \right) \sigma}{\bar{\rho}\,h^2}}\right)_x}\\{}
\dsl{\bar{\sigma}_t=}&\dsl{\left( -\frac12 \frac{(h-2\,\eta){\sigma}^{2}}{\bar\rho\,h^2}+\frac1{h}\drho g\eta \right)_x}
\end{array}
\right.
\end{equation} 
Trading the variable $\eta$ for the relative thickness $\xi=\eta_2-\eta_1$ as in Section \ref{eqm-sect} affects the Poisson tensor only by a factor of $2$, and lead us to express  
the Hamiltonian density as
\begin{equation}\label{HrB-wx}
H_{{\rm red},B}(\xi,\bar\sigma)=\frac{1}{8h\bar\rho}\left(h^2-\xi^2\right)\, {\bar\sigma}^2+\frac{g}8\drho \xi^2,
\end{equation} 
where we have used the fact that $\int_\RR \xi \, dx$ is a Casimir functional for the Darboux Poisson tensor. With the trivial rescaling 
\[
 \xi=h\xi^*
\]
we get
\begin{equation}\label{HrB-wx-star}
H_{{\rm red},B}(\xi^*,\bar{\sigma})=\frac{ h}{8 \bar\rho\,}\left(1-{\xi^*}^2\right)\, {\bar\sigma}^2+\frac{g\, h^2 }8\drho {\xi^*}^2.
\end{equation} 
Introducing the so-called reduced gravity constant
\begin{equation}
 \label{gtilde}
 \widetilde{g}:= g\,\drho, 
\end{equation} 
which, in view of the double scaling (\ref{douscalim}) is a finite quantity, and performing the additional  scaling
\begin{equation}\label{Bou-newscaling}
 \bar\sigma=\sqrt{\bar\rho h} \, \sigma^*,
\end{equation} 
we get the Boussinesq Hamiltonian in the final form
\begin{equation}\label{HrB-phy}
 H_{{\rm red},B}(\xi^*,\sigma^*)=\frac{h^2}8 \left((1-{\xi^*}^2)\, {\sigma^*}^2+\widetilde{g} {\xi^*}^2\right).
\end{equation} 


In the next section we will discuss in great detail the Boussinesq limit. However, we remark here that the change of variables and the scalings we made so far 
are convenient even without assuming the Boussinesq approximation for the kinetic term. Indeed, if we write the full Hamiltonian density appearing in~(\ref{Hred}) in terms of the variables $(\xi^*, \sigma^*)$, taking into account that equation~(\ref{psidef}) now reads
$
\varphi=h\bar\rho(1-\frac{\drho}{2\bar\rho}\, \xi^*)\, , 
$ we get
\begin{equation}\label{Hdef}
H_{\rm red}(\xi^*,\sigma^*)=\frac18 {g\drho h^2}\left((1-{\xi^*}^2)\, \frac{{\sigma^*}^2}{1-r\, \xi^*}+{\xi^*}^2\right), 
\end{equation} 
with
\[
r=\frac{\drho}{2\bar\rho}.
\]
Hence, in the double scaling limit it appears that this Hamiltonian can be viewed as a  deformation of $H_{{\rm red},B}$, with respect to the ``small" parameter $r$.

\section{Integrability of the Boussineq limit}\label{B-limit}
From now on we fix a suitable non-dimensional version of the Hamiltonian picture of the previous section by taking 
\begin{equation}
\label{Hdef1}
 J=-\left( \begin{array}{cc}
            0 & \partial \\ \partial & 0
           \end{array}
 \right), \qquad H(\xi,\sigma)= \frac{1}{4}\left(\frac{(1-\xi^2)\sigma^2}{1-r\xi}+ \xi^2\right)  
\end{equation}
where we also have dropped all the asterisks from non-dimensional variables. 
In this section we shall explicitly discuss some known  facts about the integrability of the 
dispersionless two-fluid system when the Boussinesq approximation is applied, with an eye towards  the further developments to be discussed in Section~\ref{sect-defo}.

In this framework the  Boussinesq approximation coincides with the zero-th order term in the formal expansion in $r$ of the Hamiltonian of equation~(\ref{Hdef1}). 
Explicitly, the Hamiltonian for the Boussinesq approximation is 
\begin{equation}\label{BH}
H_{0}(\xi,\sigma)=\frac14\left((1-{\xi}^2)\, {\sigma}^2+{\xi}^2\right).
\end{equation} 
As recalled in Section \ref{KH},
the ensuing Hamiltonian equations of motion can be written as
\begin{equation}\label{B-eqmH}
\left( 
\begin{array}{c}\medskip
\xi_t\\
\sigma_t
\end{array}
\right)+\left( 
\begin{array}{c}\medskip
\left(\frac{\partial H_{0}}{\partial \sigma}\right)_x\\
\left(\frac{\partial H_{0}}{\partial \xi}\right)_x
\end{array}
\right)
=\left(\begin{array}{c} 0\\0\end{array}\right)\, , 
\end{equation}
that is, explicitly,
\begin{equation}\label{B-eqmex}\left\{
\begin{array}{c}\medskip
\xi_t+\frac12\left((1-\xi^2)\sigma\right)_x=0\>\\
\sigma_t+\frac12\left((1-\sigma^2)\xi\right)_x=0\, .
\end{array}\right.
 \end{equation} 
The quasilinear form of this system is
\begin{equation}\label{B-ql}
\left( 
\begin{array}{c}\medskip
\xi_t\\
\sigma_t
\end{array}
\right)+
 \left(\begin {array}{cc} \xi\,\sigma&\frac12\,(
 {\xi}^{2}-1)
\\ \noalign{\medskip}\frac12\,({\sigma}^{2}-1)&\xi\,\sigma\end {array}
 \right)
\left( 
\begin{array}{c}\medskip
\xi_x\\
\sigma_z
\end{array}
\right)=\left(\begin{array}{c}\medskip 0\\0\end{array}\right).
 \end{equation} 
Characteristic velocities and Riemann invariants can be  obtained  from this representation (following \cite{MTTRM04, BM12}) as
\begin{equation}\label{cvri} 
 \lambda_\pm=\xi\sigma\pm\frac12\sqrt{(1-\xi^2)(1-\sigma^2)}.\quad R_\pm =\mp \xi\sigma+\sqrt{(1-\xi^2)(1-\sigma^2)}\,.
\end{equation} 
Thus, the relation between characteristic velocities and Riemann invariants (to be further discussed in Section \ref{Godograph}), 
turns out to be 
\begin{equation}\label{cv-Ri}
\lambda_+=\frac14(3 R_- -R_+) \, , \quad \lambda_-=\frac14(R_+-3R_-).
\end{equation} 
These are exactly the relations between characteristic velocities and Riemann invariants of the Airy system, also known as dispersionless defocusing non-linear Schr\"odinger (\dNLS) equation, 
written in the so-called Madelung variables that are obtained by parameterizing the Schr\"odinger wave function as
\begin{equation}\label{nls}
 \psi=u^2\exp \left(i \int \hspace{-0.1cm} v\,  \d x \right)\, . 
\end{equation} 
The resulting system is, explicitly,
\begin{equation}\label{dNLSE}
 u_t+(u v)_x=0, \qquad v_t+vv_x+u_x=0.
\end{equation}
The coordinate change that sends our Boussinesq system into the \dNLS\ system can be deduced from the structure of the characteristic velocities as
\begin{equation}
\label{CoV}
 u=(1-\xi^2)(1-\sigma^2), \qquad v= 2 \xi\sigma.  
\end{equation}
What is most important for us is that this  system, besides being integrable via the hodograph method, is well known to admit 
infinite families of constants of the motion. In the next section we shall recall how these are constructed.
\subsection{Some properties of the \dNLS\  equations}
It is well known that the \dNLS\ (or Airy)  system (\ref{dNLSE}) with 
can be obtained via local, compatible Poisson 
structures
\begin{equation}
\label{P0P1airy}
 P_0=\left( 
 \begin{array}{cc}
  0&\partial \\
  \partial & 0
 \end{array}
 \right) \qquad
 P_1=\left( 
 \begin{array}{cc}
  u \partial + \partial u  & v \partial \\
  \partial v & 2 \partial
 \end{array}
 \right)
 \end{equation}
 with appropriate definitions of Hamiltonian functionals. 
 A perhaps lesser known property of the \dNLS\ equations is that they admit a third {\em local} Poisson structure~(see, e.g., \cite{kuper})
  \begin{equation}\label{3-poi}
 P_2=P_1(P_0)^{-1}P_1=\left( 
 \begin{array}{cc}
  2u v \partial + 2\partial u v  & 2u \partial + 2\partial u  +v^2 \partial   \\
  2u \partial + 2\partial u  + \partial v^2  & 2  v \partial + 2 \partial v
 \end{array}
 \right),
\end{equation}
obtained via the recursion tensor
\begin{equation}\label{N-rec}
 N:=P_1\cdot P_0^{-1}=\left(
 \begin{array}{cc}  v& u+\partial u \partial^{-1}\\ 2& \partial v \partial^{-1}
 \end{array}\right).
\end{equation}
This iterated Poisson structure is important in our context,  thus we review the well-known 
multi-Hamiltonian representation of the \dNLS/Airy system taking $P_2$ into account next. 

An infinite number of conserved quantities for the \dNLS\ system is encoded 
in the generator 
\begin{equation}
 Q(\lambda)=-\frac14 \sqrt{(v-\lambda)^2-4u},
\end{equation}
obtained  by solving the system
\begin{equation}
 (P_1-\lambda P_0)\cdot \left(\begin{array}{c}Q_u\\Q_v\end{array}\right)=0.
\end{equation}
in the ring of formal power series in $\la$. Thus, $Q(\la)$ is  the formal generator of the Casimir of the Poisson pencil $P_1-\la P_0$.

Expanding this generator around $\la=\infty$ we get the family of conserved densities
\begin{equation}\label{AlgCas}\begin{split}
K(\la)=&  Q(\la) \vert_{\la\to \infty}-\frac{\lambda}{4} = \frac{v}{4}+\frac{ u}{2\lambda }+\frac{ u \, v}{2\lambda ^2}+\frac{ u \left(u+v^2\right)}{2\lambda ^3}+\\ &
\frac{ u \, v \left(3
   u+v^2\right)}{2\lambda ^4}+ \frac{ u \left(2 u^2+6 u v^2+v^4\right)}{2\lambda ^5}+\dots=\sum_{j=0}^\infty {1\over \la^j}K_j ,\end{split}
\end{equation}
to be referred to as the family of {\em polynomial} invariants. Notice that the coefficient of $\la^{-3}$ in the expansion~(\ref{AlgCas}) is a multiple of the ``standard" Hamiltonian of the \dNLS\ system, in the sense that  system~(\ref{dNLSE})  can be written as 
\begin{equation}\label{cms}
\left(\begin{array}{l} u\\v\end{array}\right)_t=P_0 \cdot 
\left(\begin{array}{l} \delta_u\oper \\
\delta_v \oper \end{array}\right),
\quad \text{with } \oper=-\frac14 \int K_3 \d\, x= -\frac12 \int  u \left(u+v^2\right) \d x \, .
\end{equation}
(Hereafter, integration is undertood in a formal sense, and we will omit the range of integration of the conserved densities in integrals; in all cases, the dependent variables $(u,v)$  and $\xi, \sigma)$ will be assumed to be defined up to appropriate constants in order to lead to integrable conserved densities.) 

 The quantities $\mc{K}_j=\int K_j dx$ satisfy the following recursion relations:
 \begin{equation}\label{relrec}
P_0 \cdot \left(\begin{array}{l} \delta_u\mc{K}_j \\
\delta_v \mc{K}_j \end{array}\right)\> =\>P_1 \cdot \left(\begin{array}{l} \delta_u\mc{K}_{j+1} \\
\delta_v \mc{K}_{j+1} \end{array}\right)\> =\>P_2 \cdot  \left(\begin{array}{l} \delta_u\mc{K}_{j+2} \\
\delta_v \mc{K}_{j+2} \end{array}\right)\quad .
\end{equation}
In particular, this implies that the conserved integrals are in involution with respect to  any of the Poisson tensors $P_k, k=0,1,2$, and so, 
thanks to well-known properties of the Lenard-Magri chains,  with respect to  any of their linear combinations. 

The following proposition, whose proof is a matter of a straightforward computation in differential geometry, is however 
crucial for our purposes.
\begin{prop}\label{newpropo}
Under the coordinate change
\begin{equation}
\label{CoV2}
 u=(1-\xi^2)(1-\sigma^2), \qquad v= 2 \xi\sigma\, ,
\end{equation}
that identifies the Boussinesq limit of the $2$-layer fluid system with the \dNLS\  system, the Darboux Poisson structure of the 
former is sent into the linear combination $4 P_0-P_2$ of the latter. Moreover the structure $P_0$ is sent in
\begin{equation}
 P_0=\left( \begin{array}{cc}
              (P_0)^{11} 
             & 
              (P_0)^{12} 
             \\
              (P_0)^{21} 
             & 
           (P_0)^{22}  
            \end{array}
\right)
\end{equation}
where
\begin{equation}
 \begin{split}
  (P_0)^{11} =&  \left(\frac{\xi \sigma  \left(1-\xi ^2\right) }{4 \left(\xi ^2-\sigma ^2\right)^2} \right)\partial+
             \partial \left(\frac{\xi \sigma  \left(1-\xi ^2\right)  }{4 \left(\xi ^2-\sigma ^2\right)^2} \right)\\
  (P_0)^{12}=& \left(\frac{2 \xi ^2 \sigma ^2-\xi ^2-\sigma ^2}{8 \left(\xi ^2-\sigma ^2\right)^2}\right)\partial+
            \partial \left(\frac{2 \xi ^2 \sigma ^2-\xi ^2-\sigma ^2}{8 \left(\xi ^2-\sigma ^2\right)^2}\right)-
              \frac{\xi  \sigma  (\sigma  \xi_x -\xi  \sigma_x )}{4 \left(\xi ^2-\sigma ^2\right)^2}\\
   (P_0)^{21}=& \left(\frac{2 \xi ^2 \sigma ^2-\xi ^2-\sigma ^2}{8 \left(\xi ^2-\sigma ^2\right)^2}\right)\partial+
             \partial \left(\frac{2 \xi ^2 \sigma ^2-\xi ^2-\sigma ^2}{8 \left(\xi ^2-\sigma ^2\right)^2}\right)+
             \frac{\xi  \sigma  (\sigma  \xi_x -\xi  \sigma_x )}{4 \left(\xi ^2-\sigma ^2\right)^2}\\
   (P_0)^{22}=& \left( \frac{\xi \sigma \left(1 -\sigma ^2\right)}{4 \left(\xi ^2-\sigma ^2\right)^2} \right) \partial 
              +\partial \left( \frac{\xi \sigma \left(1 -\sigma ^2\right)}{4 \left(\xi ^2-\sigma ^2\right)^2} \right)
 \end{split}
\end{equation}
\end{prop}
In Appendix \ref{app-P1} we report also  the structure of $P_1$ (\ref{P0P1airy}) in coordinates $(\xi,\sigma)$. \\
This simple fact proves a  Liouville-like integrability  of the Boussinesq limit of the $2$-layer fluid system. Indeed, all the quantities obtained by the bi-Hamiltonian recursion  we recalled above are, once written  in the proper coordinates $(\xi, \sigma)$, conserved quantities for the $2$-layer Boussinesq system. The  Boussinesq 
Hamiltonian corresponds to  \dNLS\  Casimir density $-\dsl{{u}/{4}}$, i.e., a multiple of the coefficient of $\la^{-1}$ in the expansion (\ref{AlgCas}). 

\begin{rem}
For the sake of completeness, we notice that two more infinite sequences  of (mutually commuting) conserved quantities can be obtained for the \dNLS/Airy system. 
The first one 
is obtained by formally Taylor-expanding $Q(\lambda)$ around 
$\lambda=0$.  The family of {\em algebraic} conserved quantities
\begin{equation}\begin{split}
 &Q(\la)\vert_{\la\to 0}=-\frac{1}{4} \sqrt{v^2-4 u}+\frac{\lambda  v}{4 \sqrt{v^2-4
   u}}+\frac{\lambda ^2 u}{2 \left(v^2-4
   u\right)^{3/2}}+\frac{\lambda ^3 u v}{2 \left(v^2-4
   u\right)^{5/2}}\\&+\frac{\lambda ^4 u \left(u+v^2\right)}{2
   \left(v^2-4 u\right)^{7/2}}+\frac{\lambda ^5 u v \left(3
   u+v^2\right)}{2 \left(v^2-4 u\right)^{9/2}}+O\left(\lambda
   ^6\right)
   \end{split}
\end{equation}
is thus obtained.
These conserved densities in $\xi,\sigma$ physical  variables become 
\begin{equation}
\begin{split}
&Q(\la)\vert_{\la\to 0}=-\frac{1}{2} \sqrt{\xi ^2+\sigma ^2-1}+\frac{\lambda  \xi  \sigma
   }{4 \sqrt{\xi ^2+\sigma ^2-1}}+\frac{\lambda ^2 \left(\xi
   ^2-1\right) \left(\sigma ^2-1\right)}{16 \left(\xi ^2+\sigma
   ^2-1\right)^{3/2}}\\&+\frac{\lambda ^3 \xi  \left(\xi ^2-1\right)
   \sigma  \left(\sigma ^2-1\right)}{32 \left(\xi ^2+\sigma
   ^2-1\right)^{5/2}}+\frac{\lambda ^4 \left(\xi ^2-1\right)
   \left(\sigma ^2-1\right) \left(\xi ^2 \left(5 \sigma
   ^2-1\right)-\sigma ^2+1\right)}{256 \left(\xi ^2+\sigma
   ^2-1\right)^{7/2}}\\&+\frac{\lambda ^5 \xi  \left(\xi ^2-1\right)
   \sigma  \left(\sigma ^2-1\right) \left(\xi ^2 \left(7 \sigma
   ^2-3\right)-3 \sigma ^2+3\right)}{512 \left(\xi ^2+\sigma
   ^2-1\right)^{9/2}}+O\left(\lambda ^6\right)
\end{split}
   \end{equation}

The second sequence of conserved densities is obtained using the well known fact that the (dispersionless) Toda equations are symmetries of the \dNLS\ equations.
In particular, the corresponding Toda Hamitonian  
\begin{equation}
S_1=u(\ln(u)-1)+\frac{u^2}{2} \, 
\end{equation}
 is conserved along the \dNLS\ flow (\ref{dNLSE}). Since $N^* dS_1=d S_2$ (where $N^*$  denotes the usual definition of the adjoint of the recursion operator $N$ in~(\ref{N-rec})),  
and 
$$
S_2= v u +v u  \log (u )+\frac{1}{6} v^3 \, , 
$$ 
iterating with $S_1$ 
yields another infinite family of conserved, mutually commuting quantities $S_n$, $n>1$. The first  elements of this additional family can be computed to be
\begin{equation}
 \label{Todaintegrals}
 \begin{split}
  S_3=& 2 v^2 u +\left(v^2 u +u ^2\right) \log (u )+\frac{1}{12} v^4+\frac{1}{2} u ^2\\
  S_4=& \frac{1}{60} v \left(160 v^2 u +3 v^4+210 u ^2\right)+\frac{1}{60} v \left(60 v^2 u +180 u ^2\right) \log
   (u )
   \, . 
\end{split}
\end{equation}
The coresponding conserved quantities in terms of the physical  variables  $\xi,\sigma$ are
\begin{equation}
\begin{split}
S_1=&\frac{1}{2} \left(4 \xi ^2 \sigma ^2-2 \left(1-\xi ^2\right) \left(1-\sigma ^2\right)\right)+\left(1-\xi ^2\right) \left(1-\sigma
   ^2\right) \log \left(\left(1-\xi ^2\right) \left(1-\sigma ^2\right)\right)\\
S_2=& \frac{4 \xi ^3 \sigma ^3}{3}+2 \xi  \left(1-\xi ^2\right) \left(1-\sigma ^2\right) \sigma +2 \xi  \left(1-\xi ^2\right)
   \left(1-\sigma ^2\right) \sigma  \log \left(\left(1-\xi ^2\right) \left(1-\sigma ^2\right)\right)\\
S_3=& \left(1-\xi ^2\right) \left(1-\sigma ^2\right) \left(4 \xi ^2 \sigma ^2+\left(1-\xi ^2\right) \left(1-\sigma ^2\right)\right) \log
   \left(\left(1-\xi ^2\right) \left(1-\sigma ^2\right)\right)\\&+\frac{1}{12} \left(16 \xi ^4 \sigma ^4+96 \xi ^2 \left(1-\xi
   ^2\right) \left(1-\sigma ^2\right) \sigma ^2+6 \left(1-\xi ^2\right)^2 \left(1-\sigma ^2\right)^2\right)\\
S_4=& \frac{8 \xi ^5 \sigma ^5}{5}+7 \xi  \left(1-\xi ^2\right)^2 \left(1-\sigma ^2\right)^2 \sigma 
+\frac{64}{3} \xi ^3 \left(1-\xi ^2\right) \left(1-\sigma ^2\right)
   \sigma ^3\\&+2 \xi  \left(1-\xi ^2\right)
   \left(1-\sigma ^2\right) \sigma  \left(4 \xi ^2 \sigma ^2+3 \left(1-\xi ^2\right) \left(1-\sigma ^2\right)\right) \log
   \left(\left(1-\xi ^2\right) \left(1-\sigma ^2\right)\right)
   \, . 
\end{split}
\end{equation}

\end{rem}

\section{The $r$-expansion for non-Boussinesq deformations}\label{sect-defo}
Let us now go back to the study of the two-layer equations of motion (\ref{Hdef1}), 
and in particular to the task of finding conserved quantities for this system,
 written in the quasilinear form 
\begin{equation}\label{eqhr=1}
\xi_t=-(H_{ \sigma})_x,\qquad \sigma_t=-(H_{ \xi})_x,
\end{equation} 
with 
\[
H(\xi,\sigma)= \frac{1}{4}\left(\frac{(1-\xi^2)\sigma^2}{1-r\xi}+ \xi^2\right)  \, . 
\]
We have discussed the integrability of the Boussinesq limit $r=0$ of these equations. 
One could ask whether the  generic $r\neq0$ case is similarly integrable. 
We have not succeeded in finding a second Hamiltonian structure for this case, however 
it can be proved that if one existed it would not correspond to a linear combination of 
the Poisson structures induced by $P_j$, $j=0,1,2$. Hence, the construction of conserved quantities for the deformed system cannot proceed as in the usual framework which we used for the Boussinesq case. 

In general, 
for one-dimensional Hamiltonian systems in Darboux coordinates,  conserved quantities are functionals 
$\mathcal{F}$ satisfying the commutation relation
\begin{equation}
 \{\mathcal{F},\mathcal{H}\}\equiv \int_{\mathbb R} dx  \big(F_\xi (H_\sigma)_x +F_\sigma (H_\xi)_x \big)= 0.
\end{equation}
The functional $\mc{F}$ is a conserved density only if the integrand 
$(F_\xi (H_\sigma)_x +F_\sigma (H_\xi)_x )$
is a total spatial derivative (we are assuming  that  both fields satisfy appropriate boundary conditions, e.g., vanish together with their spatial derivatives  as $|x|\to \infty$ for integrals of the whole real line). 
A direct computation shows that 
this property translates into the equation
\begin{equation}
\label{CQcon}
F_{\xi \xi} H_{\sigma \sigma}=H_{\xi \xi} F_{\sigma \sigma} 
\end{equation}
for the densities.
In our case this yields the following PDE for the density $F$:
\begin{equation}
\label{Feqtot}
\left( {\xi}^{2}-1 \right) {\frac {\partial ^{2}}{\partial {\xi}^{2}}
}F \left( \xi,\sigma \right)={\frac {{r}^{3}{\xi}^{3}-{r}^{2}{\sigma}^{2}-3\,{r}^{2}{\xi}^{2}+3\,r
\xi+{\sigma}^{2}-1}{ \left( 1- r\xi\right) ^{2}}}\, {\frac {\partial ^{2}}{\partial {\sigma}^{2}}
}F \left( \xi,\sigma \right).
\end{equation}
Explicit solutions of (\ref{CQcon})-(\ref{Feqtot}) are in general not available, hence we turn to the perturbative analysis in the small $r$-limit expansion of the Hamiltonian 
$H(\xi,\sigma)$. The relevance of this limit lies in the physical significance of the normalized density difference $r$ in the original model, since in most applications (e.g., in fresh--salted water systems or in meteorology-related problems of dry--wet air)  naturally occurring  density variations are invariably small when appropriately normalized. 
%

The first order deformation  in the small parameter $r$ is
 \begin{equation}\label{H-r-1}
H(\xi,\sigma)=H_0(\xi,\sigma)+r H_1(\xi, \sigma)+o(r)=
\frac14 \left((1-\xi^2)\, {\sigma}^2+
\xi^2\right)+
r\, \frac14 \xi (1-{\xi}^2)\sigma^2+o(r),
\end{equation} 
and, correspondingly, the deformed Hamiltonian equations of motion, 
read explicitly 
\begin{equation}
\label{eqrhex}\begin{split}
\xi_t=&\frac12\Big(\sigma
\,{\xi}^{2}-\,\sigma+r\, (\sigma\,{\xi}^{3}-\,\sigma\,\xi ) \Big)_x+o(r)\\
\sigma_t=&\frac12\Big(
\xi\,{\sigma}^{2}-\,\xi
+r \big({\xi}^{2}{\sigma}^{2}+{1 \over 8}\, ( {\xi}^{2}-1 ) {
\sigma}^{2} \big) 
\Big)_x+o(r)\end{split}
\end{equation}
Although these equations, and in particular the associated initial value problem, can in principle be studied with standard Riemann methods, these turn out to be quite cumbersome, even if explicit expressions of the Riemann invariants can be computed
(we shall come back to this issue in  Section \ref{Godograph}). 
Instead, we turn to the task of defining conserved quantities of these first-order  deformed equation.

According to our perturbative approach, we substitute $H=H_0+r\, H_1\equiv H_{r,1}$ (defined by equation~(\ref{H-r-1})) 
and seek an approximate constant of the motion $F$ in the form
\begin{equation}
\label{F-r}
F_{r,1}=F_0+ r F_1
\end{equation}
satisfying equation (\ref{CQcon}) at first order in $r$.   This reduces to
\begin{equation}\label{CQcon-1}
{F_{r,1}}_{\, \xi \xi} {H_{r,1}}_{ \sigma \sigma}={H_{ r,1}}_{\,\xi \xi} {F_{ r,1}}_{\sigma \sigma} +O(r^2). 
\end{equation}
Dropping terms of order $O(r^2)$ or higher we get, explicitly, 
\begin{equation}\label{CQcon-1-ex}
F_{0\, \xi \xi} H_{0\, \sigma \sigma}+r\left(F_{1\, \xi \xi} H_{0\, \sigma \sigma}+F_{0\, \xi \xi} H_{1\, \sigma \sigma}\right) =H_{0\,\xi \xi} F_{0\,\sigma \sigma}+r\left(H_{1\,\xi \xi} F_{0\,\sigma \sigma}+H_{0\,\xi \xi} F_{1\,\sigma \sigma}\right).
\end{equation}
Of course, at leading order we have to require that $F_0$ be the density of a conserved quantity for the Boussinesq limit (and hence we have plenty of such solutions, as described in Section \ref{B-limit}). 
At order $O(r)$ we have 
\begin{equation}
\label{CQcon-1-ex1}
F_{1\, \xi \xi} H_{0\, \sigma \sigma}+F_{0\, \xi \xi} H_{1\, \sigma \sigma} =H_{1\,\xi \xi} F_{0\,\sigma \sigma}+H_{0\,\xi \xi} F_{1\,\sigma \sigma},
\end{equation}
i.e., substituting the expression of $H_{r,1}$, 
\begin{equation}\label{eqF1}\begin{split}
\left(1 -{\sigma}^{2} \right)& {\frac {\partial ^{2}}{\partial {
\sigma}^{2}}}F_{{1}} \left( \xi,\sigma \right) - \left( 1-{\xi}^{2} \right) {\frac {\partial ^{2}}{
\partial {\xi}^{2}}}F_{{1}} \left( \xi,\sigma \right)  = \\  & 3\,\xi\,{\sigma}^{2}{
\frac {\partial ^{2}}{\partial {\sigma}^{2}}}F_{{0}} \left( \xi,\sigma
 \right) - \left( 1-{\xi}^{
2} \right) \xi\,{\frac {\partial ^{2}}{\partial {\xi}^{2}}}F_{{0}}
 \left( \xi,\sigma \right) \end{split}
\end{equation}
The problem of finding suitable deformations of the conserved quantities of the Boussinesq limit thus reduces to the problem of finding solutions to this inhomogeneous linear equation for $F_1(\xi,\sigma)$. As we shall see below, when $F_0$ is a polynomial constant of motion for the Boussineq system,  this equation can be solved explicitly within the class of polynomials.

\subsection{Deformation of the polynomial conserved quantities}\label{defpolconst}
The polynomial constants of motion for the Boussinesq limit are
 those obtained expanding the generator 
\begin{equation}
Q(\la)=-\frac14 \sqrt { \left( \lambda-v \right) ^{2}-4\,u}
\end{equation}
around $\la=\infty$. Under the substitution (\ref{CoV}), that is,
\[
 u=(1-\xi^2)(1-\sigma^2), \qquad v= 2 \xi\sigma.  
\] 
this generator becomes
\begin{equation}\label{Polgenxi}
\tilde{Q}(\la)=-\frac14\sqrt {{\lambda}^{2}-4\, \xi\sigma\,\lambda+4(\,{\sigma}^{2}+\,{\xi}^{2}-1) },
\end{equation}
and the first few densities are explicitly
\begin{equation}\label{Cex}
\begin{split}
F_{0,1}&=\frac12\,\xi\,\sigma\\
F_{0,2}&=\frac12{ { (1-\xi^2)(1-\sigma^2) }}\\
F_{0,3}&= {\xi\,\sigma\,(1-\xi^2)(1-\sigma^2 )}\\
F_{0,4}&=\frac12 { (1-\xi^2)(1-\sigma^2)  \left( 5\,{\xi}^{2}{\sigma}^{2}-{
\sigma}^{2}-{\xi}^{2}+1 \right) }\\ 
F_{0,5}&= {\xi\,
\sigma\,(1-\xi^2)(1-\sigma^2)  \left( 7\,{\xi}^{2}{\sigma}^{2}-3\,
{\sigma}^{2}-3\,{\xi}^{2}+3 \right) }+\\ 
F_{0,6}&= (1-\xi^2)(1-\sigma^2)   \left( 21\,{\sigma}^{4}{\xi}^{4}-
14\,{\sigma}^{4}{\xi}^{2}-14\,{\sigma}^{2}{\xi}^{4}+\right. \\&\quad\>\,\left. {\sigma}^{4}+16\,{
\xi}^{2}{\sigma}^{2}+{\xi}^{4}-2\,{\sigma}^{2}-2\,{\xi}^{2}+1 \right) \,. 
\end{split}
\end{equation}
(Note that the second element of this family coincides up to a factor and a constant term with the Boussinesq Hamiltonian.)
We have the following
\begin{prop}\label{prop-defo}
 Every polynomial constant of the motion $F_{0,j}$ admits a (polynomial) first order deformation $F_{1,j}$, i.e, for all integer $j$'s the quantities
 $F_{0,j}+r F_{1,j}$ satisfy equation (\ref{CQcon-1-ex}).
 \end{prop}
{\bf Proof.} The proof of this proposition is somewhat technical, hence we omit details here and report them in Appendix~\ref{a1}.

It turns out that  $F_{0,1}\propto \xi\sigma$ remains a constant of the motion for the deformed system, so that for the deformation $F_{1,1}=0$. 
Similarly,  we already have the deformation of the Hamiltonian  $ H_{0}=\frac{1}{4}(F_{0,2}-1)$ 
as (see equation (\ref{H-r-1}))
\[
F_{1,2}=\frac14 \xi (1-{\xi}^2)\sigma^2 \, . 
\]
The first non-trivial deformations can be found to be
\begin{equation}\label{result}
\begin{split}
F_{1,3}=&\frac12\,\sigma\, \left( 4\,{\sigma}^{2}{\xi}^{4}-6\,{\sigma}^{2}{\xi}^{2}
-{\xi}^{4}+2\,{\sigma}^{2}+6\,{\xi}^{2} \right) 
\\
F_{1,4}=&\frac1{10}\,\xi\, \left( 75\,{\sigma}^{4}{\xi}^{4}-130\,{\sigma}^{4}{\xi}^{2
}-40\,{\sigma}^{2}{\xi}^{4}+55\,{\sigma}^{4}+140\,{\sigma}^{2}{\xi}^{2
}+{\xi}^{4}-100\,{\sigma}^{2}-30\,{\xi}^{2} \right) \\
F_{1,5}=&\frac12\,\sigma\, \left( 56\,{\sigma}^{4}{\xi}^{6}-110\,{\sigma}^{4}{\xi}^
{4}-45\,{\sigma}^{2}{\xi}^{6}+60\,{\sigma}^{4}{\xi}^{2}+139\,{\sigma}^
{2}{\xi}^{4} \right. \\ &\left.
+5\,{\xi}^{6}-6\,{\sigma}^{4} -111\,{\xi}^{2}{\sigma}^{2}-
41\,{\xi}^{4}+17\,{\sigma}^{2}+51\,{\xi}^{2} \right)\\
F_{1,6}=
&\frac1{35}\,\xi\, \left( 3675\,{\xi}^{6}{\sigma}^{6}-8085\,{\sigma}^{6}{\xi}
^{4}-3920\,{\sigma}^{4}{\xi}^{6}+5425\,{\sigma}^{6}{\xi}^{2}+\right. 
\\ &-11970\,{
\sigma}^{4}{\xi}^{4}+861\,{\sigma}^{2}{\xi}^{6}-1015\,{\sigma}^{6} -
10780\,{\sigma}^{4}{\xi}^{2}-4711  \,{\sigma}^{2}{\xi}^{4}+
\\&\left.-16\,{\xi}^{6}
+2730\,{\sigma}^{4}+6055\,{\xi}^{2}{\sigma}^{2}+322\,{\xi}^{4}-2205\,{
\sigma}^{2}-700\,{\xi}^{2} \right) 
\end{split}
\end{equation}
\begin{rem}
All the integrals of the motion of the  Boussinesq approximation mutually Poisson-commute among themselves. There is no {\it a priori} reason for the deformed Hamiltonian to be in involution. We have explicitly checked up to $k=28$ that the first  order deformed conserved quantities $F_{0,k}+r\,F_{1,k}$  are in involution up to terms of  order $r^2$. 
\end{rem}

\begin{rem}
Our existence results make contact with the theory of Birkhoff normal forms for Hamiltonian systems.  In the finite-dimensional case the possibility of deforming quadratic Hamiltonians up to the first order, preserving integrability is well studied and settled. In the infinite dimensional case, there are some general results in \cite{Bamba}, mostly aimed at obtaining the infinite dimensional version of the ``problem of small divisors". However, the starting point zeroth order Hamiltonian in these works is the quadratic Hamiltonian of the D'Alembert (or Klein-Gordon) wave equations. Our example seems to lie outside of such a  class, since our undeformed Hamiltonian is that of the {\em non-linear} \dNLS\ equation.
\end{rem}


\section{Perturbed Riemann invariants and a class of hodograph solutions}\label{Godograph}
It is well known (see \cite{Ts91}) that the Cauchy problem for a quasi-linear diagonalizable system of conservation laws can be in general solved by 
means of the (generalized) hodograph method. In particular,  the local hodograph form of the solutions 
of a quasilinear system assumes a particularly simple form whenever expressed 
by means of Riemann invariants. 

As already anticipated in \S\ref{intro}, the problem of computing explicitly the Riemann invariants for the full equations~(\ref{e-m-full}) with generic values of $r$ runs into 
a non-separable ODE problem, preventing explicit solutions to be found. The counterpart of the perturbative approach in the limit $r\to 0$ for the constant of motion can however be explored. Thus,  following the same strategy, we seek  first order expansions of the characteristic velocities and Riemann invariants in the form
\begin{equation}\label{expans-Rl}
 \la_\pm=\la^0_\pm+r\la^1_\pm=o(r), \qquad R_\pm=R^0_\pm+r R^1_\pm+o(r),
\end{equation} 
satisfying the corresponding Riemann system of equations
\begin{equation}\label{RE}
\left\{
\begin{array}{l}
 R_{+,t}+\la_+(R_+, R_-)R_{+, x}=0\\
 R_{-,t}+\la_-(R_+, R_-)R_{-, x}=0,
\end{array}
\right.
\end{equation} 
whose  expansions for small $r$ read
\begin{equation}\label{RE-exp}
\left\{
\begin{array}{l}
R_{+,t}^0+rR_{+,t}^1+\la_+^0(R_+^0, R_-^0) R_{+, x}^0+r (W_+R^0_{+,x}+\la_+^0(R_+^0, R_-^0)R_{+, x}^1)=o(r)\\
R_{-,t}^0+rR_{-,t}^1+\la_-^0(R_+^0, R_-^0) R_{-, x}^0+r (W_-R^0_{-,x}+\la_-^0(R_+^0, R_-^0)R_{-, x}^1)=o(r),
\end{array}
\right.
\end{equation} 
where
$W_\pm=\ddd{\la_\pm}{R_+}\big\vert_{(R_+^0, R_-^0)} R_{+}^1+\ddd{\la_\pm}{R_-}\big\vert_{(R_+^0, R_-^0)} R_{-}^1\,.$

The zero-th order term in this expansion yields the Boussinesq limit system. It has been studied and solved in \cite{BM12, MTTRM04}. Let us briefly re-derive 
and extend those results here.

In the Boussinesq domain of hyperbolicity
\begin{equation}
 -1 < \sigma <1, \qquad -1 < \xi <1
\end{equation}
a useful change of variables is 
\begin{equation}
 \xi=\sin(\theta), \, -\pi/2 < \theta < \pi/2, \qquad \sigma=\sin(\phi), \, -\pi/2 < \phi < \pi/2 \, .
\end{equation}
In these new variables the zeroth-order Riemann invariants $R_\pm^0$ can be chosen to be
\begin{equation}
 {R}^{0}_\pm= \cos(\phi \pm \theta) 
\end{equation}
and it can be checked that, at first order in $r$ one gets
\begin{equation}
 {R}^1_\pm=\frac{3}{2} \sin (\theta) \tan (\phi)+3 \sin (\theta\pm \phi) \mathrm{Arctanh} \left(\tan \left(\frac{\phi}{2}\right)\right) \mp \frac{5 \cos (\theta)}{2}.
\end{equation}
The characteristic velocities expressed in terms of the variables $\theta$ and $\phi$  are
\begin{equation}
\begin{split}
 \lambda_\pm=&\lambda^0_\pm + r \lambda^1_\pm +o(r) \\ =&\frac{1}{2} (-2 \sin(\theta)\sin(\phi) \pm \cos(\theta)\cos(\phi))\\
 +&\frac{r}{4} \left(\pm  \sin (\theta) (1-2 \tan^2(\phi)) {\cos(\theta) \cos(\phi)}+(3 \cos (2 \theta)-1) \sin (\phi)\right) +o(r)  \, .
\end{split}
 \end{equation}
In particular,  the zero-th order relation coincide with (\ref{cv-Ri}), 
 \begin{equation}
  \lambda^0_+=\frac{3}{4}{R}^0_+-\frac{1}{4}{R}^0_-, \qquad \lambda^0_-=\frac{1}{4}{R}^0_+ -\frac{3}{4}{R}^0_-\, .
 \end{equation}
This simple relation between characteristic velocities and Riemann invariants at leading order is clearly missing at the next order.
More importantly, although the first order term in the expansion (\ref{RE-exp}) can be explicitly computed (with the first order characteristics velocities 
expressed in terms of the Boussinesq-limit Riemann invariants $R^0_\pm$), the resulting 
quasilinear system of PDEs  
in the four variables $(R^0_\pm,R^1_\pm )$ cannot be diagonalized. Therefore, the standard procedure to solve the Cauchy problem for our deformed system cannot be applied.  
However, we can still find some hodograph solutions for our problem by using the first order deformations of the polynomial constants of motion described in 
Section \ref{sect-defo}, as we show next. 

A well-known results of the theory of quasilinear Hamiltonian equations (see, e.g.,
\cite{DGKM14} for a review), states that, for every   
density $F$ of a conserved quantity of a Hamiltonian quasilinear PDE 
with Hamiltonian density $H(\xi,\sigma)$, 
the functions $(\xi(x, t), \sigma(x, t))$, 
implicitly defined by the system
\begin{equation}
 \label{Du-eq}
 \left\{
 \begin{array}{rcl}
  x+t H_{\xi\sigma}&=&F_{\xi\sigma}\\
  tH_{\sigma\sigma}&=&F_{\sigma\sigma}
 \end{array}
 \right.
\end{equation} 
or equivalently by
\begin{equation}
 \label{Du-eqb}
 \left\{
 \begin{array}{rcl}
  x+t H_{\xi\sigma}&=&F_{\xi\sigma}\\
  tH_{\xi\xi}&=&F_{\xi\xi}
 \end{array}
 \right.
\end{equation} 
provide local solutions of the equations of motion. 

Since the quantities $K_l+rF_{1,l}$ found in Section (\ref{defpolconst})
are constants of motion in the small-$r$ asymptotics, we can use either of the above equations to 
construct approximate solutions (that is, solutions at first order in the $r$ expansion) to the deformed system.  

Besides being inherently local in $x$ (and of course in $t$) the set of initial conditions that can be reached using the deformations 
of the polynomial motion constants for the Boussinesq-system  is somewhat special,  since we can only treat those initial 
conditions that satisfy (\ref{Du-eq}) or \ref{Du-eqb}) with $t=0$ (e.g., we have to invert
 \begin{equation}
 \label{Du-eq-in}
 \left\{
 \begin{array}{rcl}\medskip
  x&=&F_{\xi\sigma}\big\vert_{\xi_0(x), \sigma_0(x)}\\
  0&=&F_{\xi\xi}\big\vert_{\xi_0(x), \sigma_0(x)}.
 \end{array}
 \right.
\end{equation} 

However, some interesting examples of local initial conditions can be obtained by using this approach. 
The first family of such initial conditions is obtained by setting $\sigma(x,0)=0$. 
This is physically interesting since it corresponds to a family of
vanishing velocity initial conditions. Indeed, from the definitions of $\sigma, \, \xi$ 
and 
the shear velocity $w=\ou_2(x)-\ou_1(x)$ we get (suppressing the time-dependence for ease of notation)
\[
 \sigma(x)=\bar{\rho}\, w(x) (1-r\xi(x)).
\]
Using this and the dynamical constraint (\ref{solvconstr}) yields
\begin{equation}
 \ou_1(x)=-\frac12 w(x)(1+\xi(x)), \quad \ou_2=\frac12 w(x)(1-\xi(x)), 
\end{equation} 
so that indeed the initial condition $\sigma=0$ corresponds to both fluids being initially at rest.
Even though the resulting system 
 \begin{equation}
 \label{Du-eq-0}
 \left\{
 \begin{array}{rcl}\medskip
  x&=&F_{\xi\sigma}\big\vert_{\xi_0(x), 0}\\
  0&=&F_{\xi\xi}\big\vert_{\xi_0(x), 0} \, , 
 \end{array}
 \right.
\end{equation}
for an $F$ taken to be a generic linear combination of the conserved quantities 
$F_{0,l}+rF_{1,l}$, would have a finite number of solutions, 
one can take advantage of the fact that the $\sigma$ factors throughout the densities $F_{0,2l+1}+rF_{1, 2l+1}$  (see Proposition \ref{prop-defo}), and so system~(\ref{Du-eq-0}) reduces to the single equation
\[
\dsl{x=\big(F_{0,2l+1}+rF_{1, 2l+1}\big)_{\xi\sigma}\big\vert_{\xi=\xi_0(x), \, \sigma=0}}\, , 
\] 
which can be locally inverted to yield $\xi_0(x)$.
\begin{rem}
 The local initial conditions which can be constructed starting from the Boussinesq polynomials are linear combinations 
 of polynomials of the form
 \begin{equation}
  {F_{0,2j+1}}_{\xi \sigma}(\xi,0)+r {F_{1,2j+1}}_{\xi \sigma}(\xi,0) = \kappa_{2j}(\xi) + r \,  \phi _{2j+1}(\xi) \qquad n \geq 1
 \end{equation}
 where $\kappa_i$ and $\phi_i$ are polynomials of degree $i$ whose explicit expression can be reconstructed from (\ref{Cex}) and (\ref{result}).
\end{rem}

\begin{figure}[b]
\centering
{\includegraphics[width=6cm]{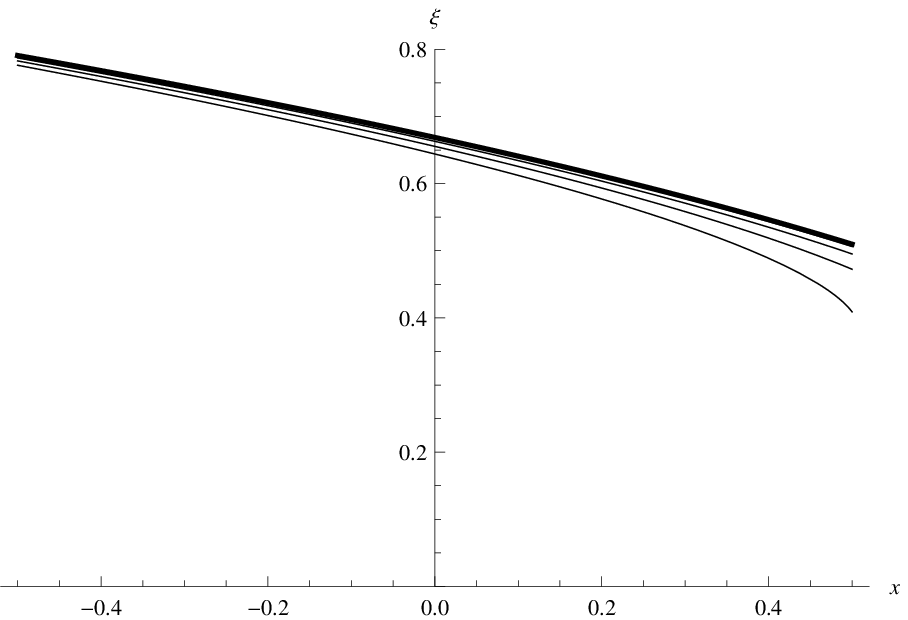}} 
\qquad \qquad
{\includegraphics[width=6cm]{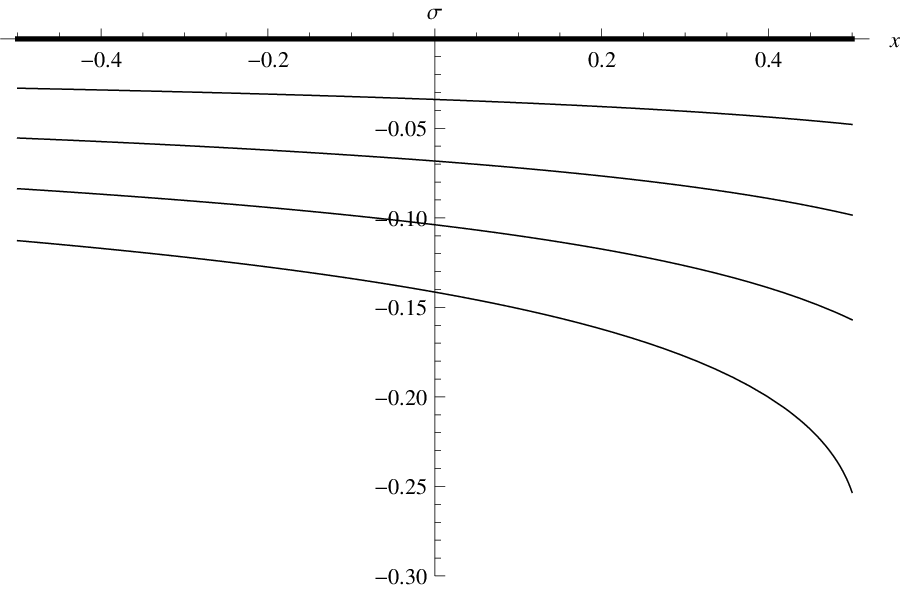}}
\caption{Evolution of the interface $\xi$ (left) and weighted vorticity $\sigma$ (right) from the initial condition constructed with the motion constant $F_{1,3}$ and zero initial $\sigma$. 
Snapshots of the local solution from non dimensional time $t=0$ to $t=2$ at $0.5$ time steps.  }
\label{fig-evo15}
\end{figure}

%
The simplest  example is the initial interface related to the conserved density 
\begin{equation}
 F_{0,3}+r F_{1,3}=\sigma\, (\xi ^3 -\xi)(\sigma ^2-1) +\frac{1}{2} \, r \sigma  \big(\xi ^4 \left(4 \sigma ^2-1\right)-6 \, \xi ^2
   (\sigma ^2-1)+2 \sigma ^2 \big)
\end{equation}
The relation
\begin{equation}
 \frac{\partial^2 \left(F_{0,3}+r \, F_{1,3}\right)}{\partial \xi \, \partial \sigma} (\xi_0,0) =  \left(-3 \xi_0 ^2
 +2r \left(3-\xi_0 ^2\right) \xi_0  +1\right) = 0
\end{equation}
gives as an initial condition  for the interface
\begin{equation}
 \xi_0(x)=\frac{\sqrt{1-x}}{ \sqrt{3}}+\frac{1}{9} \, r (x+8).
\end{equation}
which we can choose to consider, say, for  $-1/2 < x < 1/2$. The resulting evolution is depicted in Figure~\ref{fig-evo15}.

\begin{figure}
\centering
{\includegraphics[width=6cm]{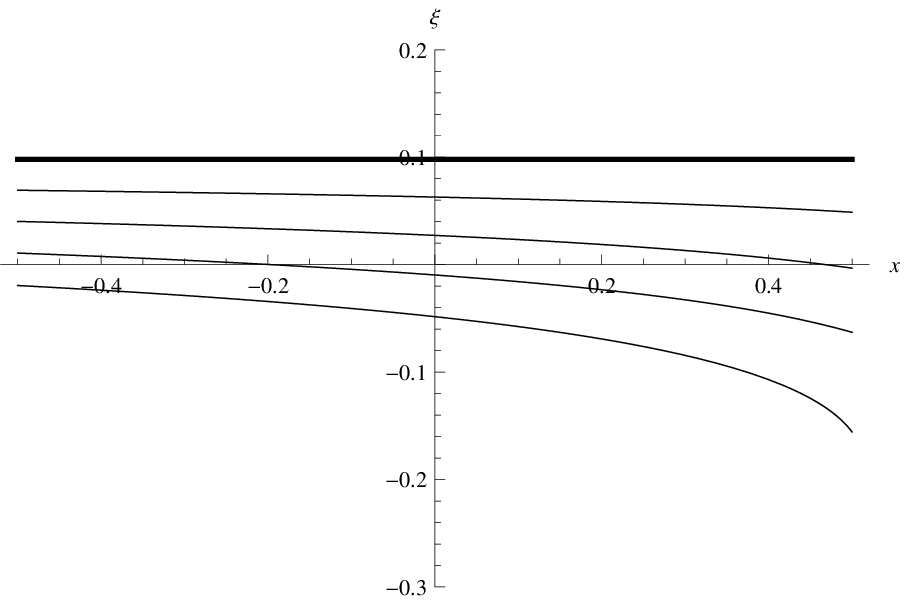}}
\qquad \qquad
{\includegraphics[width=6cm]{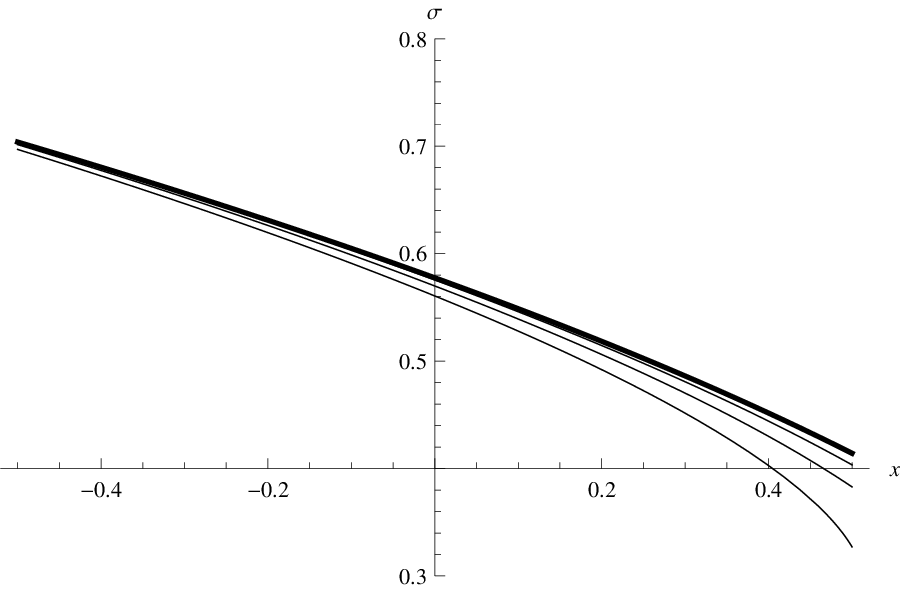}}
\caption{Same as figure~\ref{fig-evo15} but with constant initial interface. 
}
\label{fig-evo15-xk}
\end{figure}

We remark  that with the same conserved quantity $F_{0,3}+r F_{1,3}$ it is possible to study an 
initial condition for which the interface of the fluid is initially at $\xi=r+o(r)$. 
In this case the initial weighted vorticity  $\sigma$ is given by
\begin{equation}
 \sigma_0= \frac{\sqrt{1-x}}{ \sqrt{3}}+o(r)
\end{equation}
Evolution from these initial data is well defined  up to time $t=2$ in the open set $-1/2 < x < 1/2$,  as depicted by Figure~\ref{fig-evo15-xk}.

\begin{figure}[ht]
\centering
{\includegraphics[width=6cm]{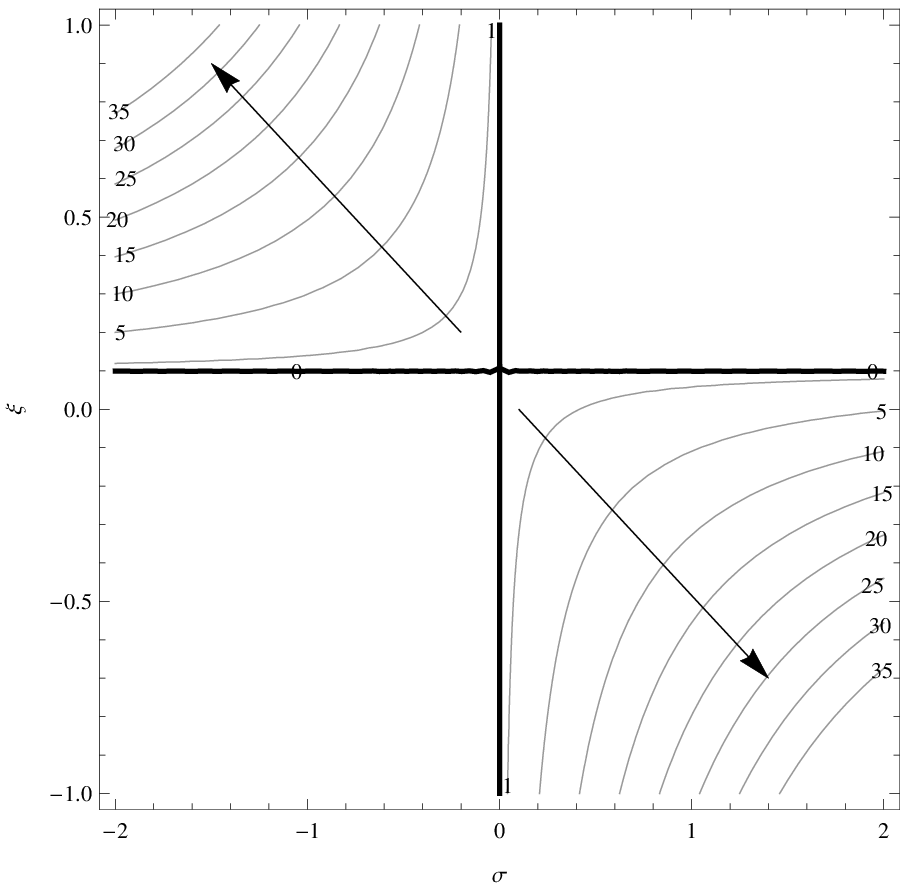}}%
\qquad \qquad
{\includegraphics[width=6cm]{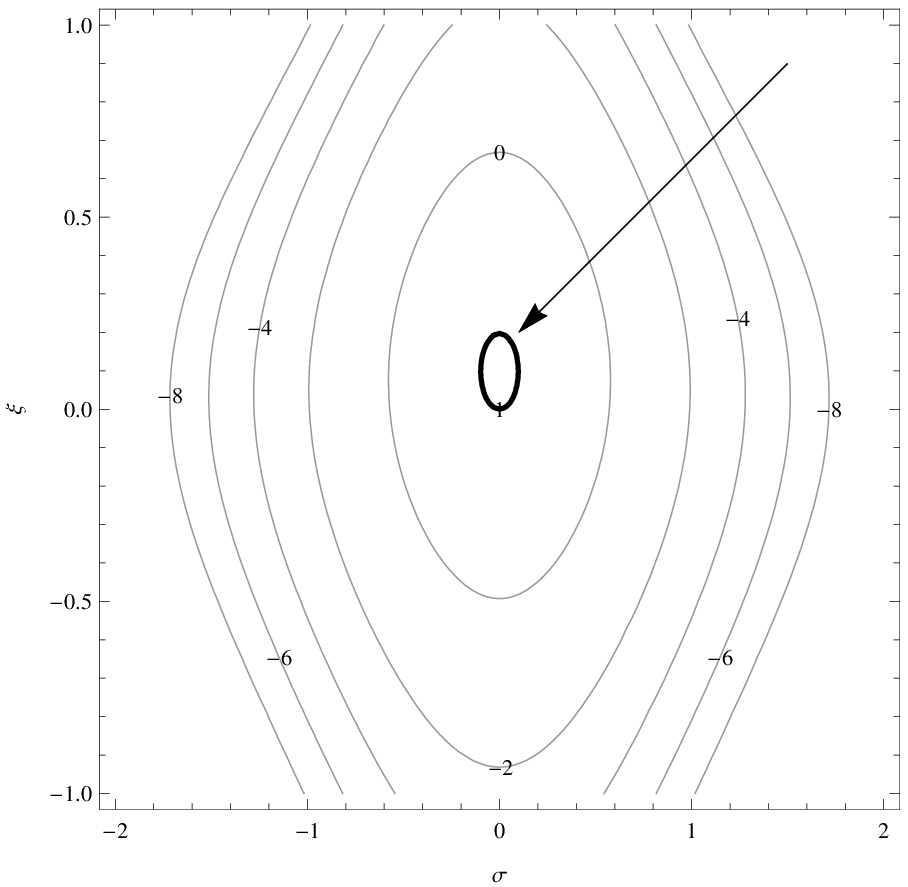}}
\caption{Evolution of the class of initial data related by the conserved quantity $F_3$ at  first order in $r$. 
Left panel: evolution in time (indicated by arrows) out of the initial data (thick line) spans the family of curves in the hodograph plane ($\xi,\sigma$).
Right panel: analogous variation in space, as the spatial coordinate $x$ grows along the direction indicated by the arrow.}
\label{fig-curveF13-tempo}
\end{figure}

Every solution constructed with the hodograph method is local because is subject to the invertibility of a map between the physical 
coordinates  $x,t$ and the fields $\xi,\sigma$ (see e.g. \cite{Ts91}). Morever the construction of the right conserved quantity $F$ associated with the
given initial condition may, in general, be as difficult a problem as the solution of the PDE system itself. To the best of our knowledge 
the reconstruction problem of the conserved quantity $F$ starting  with generic initial data is known only
for the Airy system (see \cite{TY99}). The study of a similar general result for the deformed system is interesting and will be left for future work. 
However a modicum of general information can be extracted directly from the second equation the hodograph solution (\ref{Du-eq}). This equation can 
be interpreted  as an evolution of a curve in the space $\xi,\sigma$ .
An explicit example for the initial data related to $F_{0,3}+ r F_{1,3} $ is depicted in Figure~\ref{fig-curveF13-tempo} on the left;
the family of curves in the figure is defined  by
\begin{equation}
 t= \frac{F_{\sigma \sigma}}{F_{\sigma \sigma}}
 = -24 \xi  \sigma+r \left(24 \sigma -24 \xi ^2 \sigma \right)\, + o(r) .
\end{equation}
  In a similar way, by using the linear combination of the first and second hodograph solutions 
 \begin{equation}
  x=F_{\xi \sigma} - \frac{F_{\xi \xi}}{H_{\xi \xi}}H_{\xi \sigma} 
 \end{equation}
 one obtains a family of curves parametrized by the spatial coordinate $x$ 
\begin{equation}
 x= -3 \xi ^2 \left(\sigma ^2+1\right)-3 \sigma ^2+1 + r \left( -2 \xi  \left(\xi ^2 \left(3 \sigma ^2+1\right)-3\right) \right)
  +o(r)\, . 
\end{equation} 

\section{Conclusions and discussion}
In this work, we have examined the systematic Hamiltonian reduction of Benjamin's formulation for two-dimensional stratified Euler fluids to the case of two homogeneous layers. The resulting leading order system in the long-wave approximation, with time and one spatial horizontal coordinate as independent variables, corresponds to a set of quasi-linear equations which we have framed within the theory of Hamiltonian integrable systems. In particular, we have isolated the properties of the hyperbolicity region that depend on the Hamiltonian structure, such as the tangency to sonic lines being different from simple wave tangents.  

Next, we have shown that the Boussinesq limit of negligible inertia is completely integrable as it corresponds to the dispersionless defocussing Nonlinear Schr\"odinger equation. We used the multi-Hamiltonian structure of this system to construct three infinite families of motion invariants, mutually in involution. The non-Boussinesq counterpart does not share such completely integrable structure; however, by perturbative methods based on the small inertia parameter $r$, we have shown that an infinite family of polynomial constants of motion can be constructed explicitly at leading order $O(r)$ by deformation of a specific family of the Boussinesq case. These motion invariants are then used to build examples of local solutions by the hodograph method. 

Our investigation lends itself to further generalizations, in particular: dispersion terms in the Hamiltonian reduction of Benjamin's formulation can be obtained by retaining higher order  terms in the Hamiltonian density expansion  with respect to the long-wave parameter; dispersion deformations can then be analyzed for both the Boussinesq and non-Boussinesq cases by similar Hamiltonian methods as the ones used here for the dispersionless case, and families of conserved quantities could be found, combining dispersive with non-Boussinesq deformations; in turn, these conservation laws could shed some light on the dynamics of the solution for the model systems and illustrated fundamental properties of the full Euler parent system. Study of some of these issues is ongoing and will be presented in future work.

\subsection*{Acknowledgements}
This work was carried out under the auspices of the GNFM Section
of INdAM. Partial support by NSF grants DMS-1009750, RTG DMS-0943851 and
CMG ARC-1025523, as well as by the MIUR PRIN2010-2011 project 2010JJ4KPA is
acknowledged. R.C. thanks the Department of Mathematics and its Applications of
the University   of  Milano-Bicocca for its hospitality. Discussions with B. Dubrovin are gratefully acknowledged.

\appendix
\section{Proof of Proposition \ref{prop-defo}}\label{a1}
Let us consider the family of polynomials constants of the motion for the Boussinesq approximation obtained (see equation (\ref{Cex})) by expanding the generating function 
\begin{equation}\label{Polgenxiapp}
\tilde{Q}(\la)=-\frac14\sqrt {{\lambda}^{2}-4\, \xi\sigma\,\lambda+4(\,{\sigma}^{2}+\,{\xi}^{2}-1) }
\end{equation} 
around $\la=\infty$. 
We must prove Proposition (\ref{prop-defo}), that is, 
show that each of these constants of the motion admits a polynomial deformation.

The proof of this fact can be divided in five steps.  

\paragraph{Step 1.} for the form of  $\tilde{Q}(\la)$  the following factorization properties hold:
\begin{enumerate}
\item $K_{2\,j}=(1-\xi^2)(1-\sigma^2) P_j(\xi^2, \sigma^2).$
\item $K_{2\,j+1}=\xi\sigma (1-\xi^2)(1-\sigma^2) Q_j(\xi^2, \sigma^2).$
\end{enumerate}
In particular, the most relevant property is that $K_j$'s factor through $(1-\xi^2)(1-\sigma^2)$, which is obvious since in the  Madelung variables of the \dNLS\ equation 
$Q(\la)\vert_{u=0}$ is simply $\pm\frac14 (\la-v)$. 

The finer factorization properties listed above are important for the introduction of  suitable 
subspaces on the space of bivariate polynomials in Step 3.
\paragraph{Step 2.} Since $H_1\propto \xi\sigma^2-\xi^3\sigma^2$, then
\begin{equation}\label{ternot}
H_{1,\sigma\sigma}F_{\xi\xi}-H_{1,\xi\xi}F_{\sigma\sigma}\propto 2\left( \xi(1-\xi^2)F_{\xi\xi}-3\xi\sigma^2 F_{\sigma\sigma}\right)
\end{equation}
\paragraph{Step 3}  Let $R_N$ be the subspace of polynomials generated by the monomials
\[
 \xi^{2k+1}\, \sigma^{2j}, \quad\text{with }  k,j=0\ldots N,
\]
let $S_N$ be the one generated by 
\[
  \xi^{2k}\, \sigma^{2j+1}, \quad \text{with } k=0\ldots N,\> j=0\ldots N-1
\]
and consider the operator entering the homogeneous part of equation (\ref{eqF1}):
\begin{equation}
 \label{quadra}
 \widetilde{\square}:=(1-\xi^2)\partial^2_{\xi}-(1-\sigma^2)\partial^2_\sigma.
\end{equation} 
The following holds:
\begin{enumerate}
 \item $dim(R_N)=(N+1)^2, \> dim(S_N)=N(N+1)$ 
 \item $
         \widetilde{\square}(R_N)\subset R_N\quad \widetilde{\square}(S_N)\subset S_N
       $
\item The dimension of the kernel of $ \widetilde{\square}$ restricted to $R_N$ (resp. $S_N$) is $1$. 

This means that the image of 
$ \widetilde{\square}$, seen as a map $R_N\to R_N$ (resp.  $S_N\to S_N$) is characterized by a single linear relation in $R_N$ (resp $S_N$).
\end{enumerate}
The proof of point 3 above is by direct computation, showing that the matrix representing $\widetilde{\square}$ restricted to, e.g., $R_N$ in the basis of point 2 above is 
upper triangular, with diagonal elements
\[
2\left( j+k \right)  \left( 2\,j-1-2\,k \right),
\]
whence the assertion.

\paragraph{Step 4.} An obvious observation is that, for any polynomial $P(\xi, \sigma)$ the sum of the coefficients of $\widetilde{\square}(P(\xi, \sigma))$ 
{\em vanishes}, or, in other words, 
\[
 \widetilde{\square}(P(\xi, \sigma))\big\vert_{\xi=1,\sigma=1}=0.
\]
Since $\mathrm{rk}\big(\widetilde{\square}\big\vert_{R_N}\big)=dim R_N-1$ we get that a polynomial $Q(\xi, \sigma)\in R_N$ is in the image of $\widetilde{\square}$ if and only if  
\begin{equation}\label{chacon}
Q(\xi,\sigma)\big\vert_{\xi=1, \sigma=1}=0.
\end{equation}
The same holds for $S_N$, that is, $Q\in S_N$ is in the image of $\widetilde{\square}$ if and only if  (\ref{chacon}) holds.

\paragraph{Step 5.} What is left to prove is that the LHS of equation (\ref{ternot}), i.e.
\[
 \xi(1-\xi^2)F_{\xi\xi}-3\xi\sigma^2 F_{\sigma\sigma}
\]
satisfies the characteristic condition (\ref{chacon}) when $F(\xi,\sigma)$ is one of the polynomial Hamiltonian densities. Explicitly, we have to prove that $
 \xi(1-\xi^2)H_{j, \xi\xi}-3\xi\sigma^2 H_{j, \sigma\sigma} $
satisfies (\ref{chacon}). This is immediate, since, from Step 1 we know that $H_{j}$ factors through $(1-\xi^2)$.

This ends the proof.


\section{The  limit $\mathbf{r\to 0}$}\label{naif}
\begin{figure}[htb]
\centering
\includegraphics[width=10cm, height=4.6cm]{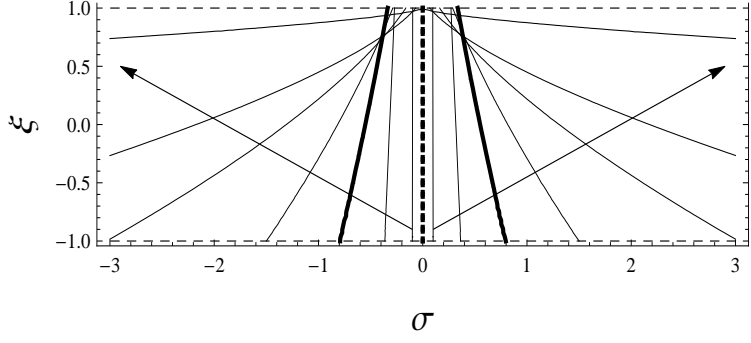}
\caption{The change of the hyperbolic region depending on $r$ 
whose bottom grows along the arrows. The dashed lines are the top and the bottom of the channel.
The vertical dotted thick line is the starting region of hyperbolicity ($r=0$). In the non-Boussinesq approximation it is of zero measure. 
The thick line corresponds to $r=(\sqrt{17}-3)/4 \simeq .280776$ which is the value of maximum amplitude of the hyperbolicity domain
at $\xi=1$. The variable $\xi$ is measured 
in unity of $h$ and $\sigma$ in unity of $\bar{\rho}\sqrt{2gh}$.}
\label{hypreg-nonB}
\end{figure}

\begin{figure}[htb]
\centering
\includegraphics[width=10cm, height=6cm]{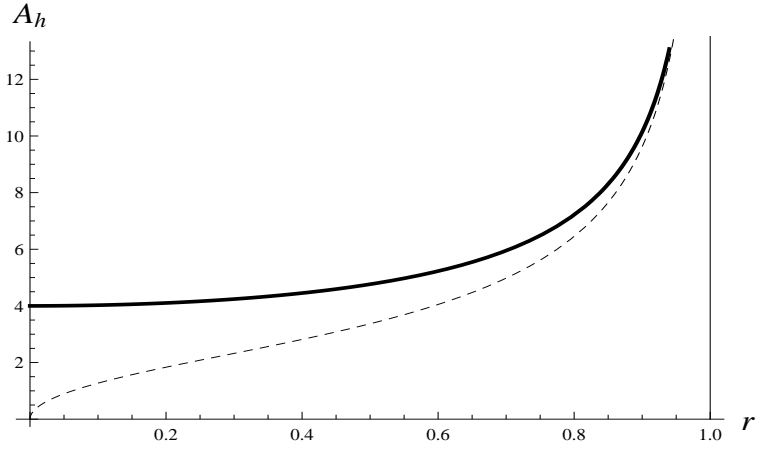}
\caption{The dashed line is the area $A_h$ measured in unity of $\bar{\rho} h \sqrt{2gh}$ of the hyperbolic region in function on $r$. 
For small $r$ the area goes to zero while goes to constant in the case of Boussinesq expansion (here 
equivalent to measure $A_h$ in unity of $\bar{\rho} h \sqrt{2\tilde{g}h}$). At $r=1$ the area diverges as $(1-r)^{-\frac12}$ in both cases.}
\label{hypreg-area-both2}
\end{figure}

In the na\"\i ve limit   in which $r\to0$  and $g$ remains finite corresponds the case of a nearly homogeneous fluid. This limit is  fundamentally different from the the Boussinesq approximation since the reduced  gravity constant $\tilde{g} = g r$ can now limit to zero. The zeroth order system in this case becomes
\begin{equation}\label{e-m-naivedrho}
\left\{
\begin{array}{cl}\medskip
\dsl{ \xi_t= }
&\dsl{- {\frac {(\left( h^2-\xi^2 \right) \sigma)_x} { 2 \bar{\rho} h }} }
 \\
\dsl{{\sigma}_t}=
&\dsl{\left( \frac{ \xi \sigma^2}{ 2\bar{\rho} h  }  \right)_x}
\end{array}
\right.
\end{equation} 
whose solutions with ${\sigma}\neq 0$ describe an ideal system of two fluids with the same density $\bar\rho$ separated by a vortex sheet. The system is 
purely elliptic, and is the prototype of a system undergoing a 
KH instability. 
Its deformation $r = o(1)$ is a small perturbation
of an elliptic system which slightly opens a hyperbolicity domain in the ($\xi, \sigma$) plane.
The hyperbolicity region is  still given by equation (\ref{hr}), i.e., 
\begin{equation}
\label{hr-app}
  |\sigma| < 
  = \bar{\rho}\sqrt{\frac{ 2 g rh  (1 - r \xi/h  )^{3}}{1-r^2}}:=\sigma_b 
  , \qquad |\xi|<h\, ,
\end{equation}
 but since $g\, r\to 0$ as $r\to 0$ the  hyperbolicity region shrinks to a tiny vertical strip around $\sigma=0$.
 The  form of the hyperbolicity domain is sketched in Figure~\ref{hypreg-nonB}.
The area $A_h$ of the hyperbolicity region  is the function  
\begin{equation}
 A_h= \rhob  \sqrt{2 g r h^3} \frac{4 \left((1+r)^{5/2}-(1-r)^{5/2}\right)}{5 r \sqrt{1-r^2}}\, ,
\end{equation}
whose graph is depicted with that of the Boussinesq approximation in Figure~\ref{hypreg-area-both2}. 
We remark that
in the limit $r\to 1$ the behavior is the same as in the $r$-expansion around the Boussinesq approximation.
\section{Poisson ${P_1}$  tensor in $(\xi,\sigma)$-coordinates}
\label{app-P1}
The Boussinesq limit of the two layer fluid admits three local Poisson structures. Two of them ($P_0$ and $P_1$ ) are already given in proposition \ref{newpropo}. 
The structure $P_1$ in physical $(\xi,\sigma)$ coordinates becomes
\begin{equation}
 P_1=\left( \begin{array}{cc}
              (P_1)^{11} 
             & 
              (P_1)^{12} 
             \\
              (P_1)^{21} 
             & 
           (P_1)^{22}  
            \end{array}
\right)
\end{equation}
where
\begin{equation}
 \begin{split}
  (P_1)^{11} =& \left( \frac{\left(1-\xi ^2\right) \left(\xi ^2+\sigma ^2\right)}{4 \left(\xi ^2-\sigma ^2\right)^2} \right) \partial 
                + \partial \left(\frac{\left(1-\xi ^2\right) \left(\xi ^2+\sigma ^2\right)}{4 \left(\xi ^2-\sigma ^2\right)^2} \right)
                  \\
  (P_1)^{12}=&  \left( \frac{\xi  \sigma  \left(\xi ^2+\sigma ^2-2\right)}{4 \left(\xi ^2-\sigma ^2\right)^2} \right) \partial 
                + \partial \left( \frac{\xi  \sigma  \left(\xi ^2+\sigma ^2-2\right)}{4 \left(\xi ^2-\sigma ^2\right)^2} \right) 
                +  \left( \frac{\left(\xi ^2+\sigma ^2\right) (\sigma  \xi_x -\xi  \sigma_x )}{4 \left(\xi ^2-\sigma ^2\right)^2} \right)
                \\
   (P_1)^{21}=& \left( \frac{\xi  \sigma  \left(\xi ^2+\sigma ^2-2\right)}{4 \left(\xi ^2-\sigma ^2\right)^2}\right) \partial 
                + \partial \left(\frac{\xi  \sigma  \left(\xi ^2+\sigma ^2-2\right)}{4 \left(\xi ^2-\sigma ^2\right)^2} \right)
                -  \left(\frac{\left(\xi ^2+\sigma ^2\right) (\sigma \xi_x   - \sigma_x  \xi )}{4 \left(\xi ^2-\sigma ^2\right)^2} \right) 
                \\
   (P_1)^{22}=& \left( \frac{\left(1-\sigma ^2\right) \left(\xi ^2+\sigma ^2\right)}{4 \left(\xi ^2-\sigma ^2\right)^2} \right) \partial 
                + \partial \left(\frac{\left(1-\sigma ^2\right) \left(\xi ^2+\sigma ^2\right)}{4 \left(\xi ^2-\sigma ^2\right)^2} \right) 
 \end{split}
\end{equation}

\end{document}